\documentclass[english,aps,pre,superscriptaddress,twocolumn,showpacs]{revtex4-1}
\usepackage{graphicx,amsfonts,amsmath,amssymb,color}
\usepackage[english]{babel}

\newcommand{\eq}[1]{$\mathrm{Eq.}$~\eqref{#1}}

\newcommand{\secref}[1]{$\mathrm{Sec.}$~\ref{#1}}

\begin{document}

\title{Weighted-ensemble Brownian dynamics simulation: Sampling of rare events in non-equilibrium systems}

\author{Justus A. Kromer}
\email{justuskr@physik.hu-berlin.de}
\affiliation{Department of Physics, Humboldt-Universit\"{a}t zu Berlin, 
Newtonstr. 15, 12489 Berlin, Germany}
\author{Lutz Schimansky-Geier}
\affiliation{Department of Physics, Humboldt-Universit\"{a}t zu Berlin, 
Newtonstr. 15, 12489 Berlin, Germany}
\author{Raul Toral} 
\affiliation{IFISC, Instituto de F{\'\i}sica Interdisciplinar y Sistemas Complejos, CSIC-UIB, E-07122 Palma de Mallorca, Spain}

\begin{abstract}
  \noindent We provide an algorithm based on weighted-ensemble (WE)
  methods, to accurately sample systems at steady state. Applying our
  method to different one- and two-dimensional models, we succeed to
  calculate steady state probabilities of order $10^{-300}$ and
  reproduce Arrhenius law for rates of order $10^{-280}$. Special
    attention is payed to the simulation of non-potential systems
    where no detailed balance assumption exists. For this large class
    of stochastic systems, the stationary probability distribution
    density is often unknown and cannot be used as preknowledge during
    the simulation.  We compare the algorithms efficiency with
  standard Brownian dynamics simulations and other WE methods.
\end{abstract}

\pacs{05.40.-a,07.05.Tp}{}
\maketitle

Rare events are ubiquitous in many biological, chemical and physical
processes \cite{moss1989noise,van2004stochastic}.  Whereas the density of states is
known in systems at thermal equilibrium, interesting phenomena often
occur in non-equilibrium systems \cite{nicolis1977self}.  Unfortunately, many such problems are
inaccessible to analytic methods. Therefore computer simulations are a
widely used tool to estimate the density of states or transition rates
between them \cite{mannella1989fast,mannella2000lecture}.  Since standard Brownian dynamic simulation
\cite{burrage1999runge,kloeden2011numerical} provides computational
costs that are inversely proportional to the state's probability,
specialized methods \cite{bhanot1987accurate,giardina2006direct,dickson2010enhanced} have
to be used to adequately sample rare events, i.e. states with low
probability or low transition rates.

In the last decades, flat histogram algorithms
\cite{wang2001efficient} have been developed, allowing one to evenly
sample states with highly different probabilities. These algorithms
are implementations of the umbrella sampling
\cite{torrie1977nonphysical}, where each state is sampled according to
a given probability distribution, the so-called umbrella
distribution. Within non-equilibrium umbrella sampling (NEUS)
\cite{warmflash2007umbrella} the space of interest is divided into
different but almost evenly sampled subregions.  The interaction
between different regions occurs solely due to probability currents
between then. Whereby the probability distribution within a region is
then calculated by performing Monte Carlo simulations.
 
In order to calculate low rates between a starting and a final state,
forward flux sampling methods can be used (for a review see)
\cite{allen2009forward}.  These methods introduce a sequence of
surfaces between these states and introduce walkers (copies of the
system) to perform weighted trajectories according to the underlying
dynamics.  If walkers cross one of the surfaces, getting closer to the
final state, new walkers with smaller weights are introduced. Finally,
many walkers with particular small weights reach the final state. The
consideration of the particular weights allows one to calculate very
low rates in a finite simulation time. Recently, extensions to these
methods have been developed to calculate both, transition rates using
umbrella sampling \cite{dickson2009separating} and probability
distributions using forward flux sampling
\cite{valeriani2009computing} algorithms.

In this work, we present an algorithm, based on the previously
developed weighted-ensemble (WE) Brownian dynamics simulations
\cite{huber1996weighted,bhatt2010steady,zhang2010weighted,bhatt2012adaptive},
that allows one to calculate the stationary probability density
function (SPDF) as well as transition rates between particular
states. Like in WE simulations the space of interest is divided into
several subregions and the probability for finding the system in them
is calculated by generating equally weighted walkers in each region.
By moving to the underlying dynamics, the walkers transport
probability between the subregions. Thus, WE methods are usually
applied to systems of Brownian particles moving in a potential
landscape \cite{bhatt2010steady,bhatt2011beyond}.  

We are
  interested in an algorithm which allows simulations of stochastic
  dynamical systems which, apriori do not obey detailed
  balance for probability fluxes or suppose some special topology of
  the flow \cite{GrHa71-2,GrHa71,Ha75,risken,LsgTol85}. Such are given for
  Brownian particles in conservative force fields under the influence
  of additive noise.  
  Even canonic dissipative systems possess a
  vanishing probability flow if transformed to the energy as dynamic
  variable \cite{Ha75,ebeling1980influence,klimo}. In consequence, both
  systems allow exact analytic solutions of the SPDF. In contrary, we
  aim to develop an algorithm which does not assume that neither the
  deterministic nor the stochastic items (see \eq{equ1DstochSystem}
  below) underly such conditions. Thus, no information on the 
  SPDF can be used for the simulations.

In general the algorithm can be applied to arbitrary dynamical systems
of the form:
\begin{eqnarray}
\label{equ1DstochSystem}
\dot{x}_n=f_n(\textbf{x})+g_n(\textbf{x}) \xi_n(t)\text{,} \ n=1,...,d \text{,}
\end{eqnarray}
where $d$ is the number of stochastic time-dependent degrees of
  freedom $x_n(t), n=1,...,d$; the functions $f_n(\textbf{x})$
describe the deterministic velocities for the $n$-th direction;
$\xi_n(t)$ represents zero-mean Gaussian noise with delta-like
correlation function
\mbox{$\langle \xi_n(t)\xi_m(t')\rangle=\delta_{nm}\delta(t-t')$}. The noise
intensity along the $n$-th direction is scaled by the functions
$g_n(\textbf{x})$, which in general depend on the vector
\mbox{$\textbf{x}=(x_1,x_2,...,x_d)^T$}. We are interested in high
precision sampling of the stationary probability current
$J_{st}(\textbf{x})$ and the SPDF $p_{st}(\textbf{x})$ of finding the
system in the $d$-dimensional cube $[x_1,x_1+dx_1],...,
  [x_d,x_d+dx_d] $ with a finite resolution. We will specify the
resolution by the number $M_{n,res}$ of evenly spaced supporting
points along the $n$-th direction, for which we will
  determine $p_{st}(\textbf{x})$.

This article is organized as follows: In section \ref{sec:algortihm}
we introduce an algorithm, based on WE methods
\cite{huber1996weighted}, that allows one to calculate low
probabilities and rates.  Afterwards, we study one- and two
dimensional model systems and analyze the algorithms efficiency
compared to Brownian dynamics simulation (BDS) and WE techniques.
\section{The algorithm} 
\label{sec:algortihm}
First, for sake of clarity, we restrict ourselves here to
particle motion in one dimension, $d=1$, under additive noise. The
  deterministic part of the dynamics \eq{equ1DstochSystem} can be
  always represented as a conservative force \mbox{$f(x)=-U'(x)$}.  The
noise strength is scaled by the parameter $D$, we put
  $g(x)=\sqrt{2 D}$.  For such systems, the SPDF is known to be
\begin{eqnarray}
\label{equ:bistablePDF}
 p_{st}(x)=Z^{-1}_{st} \ e^{-\frac{U(x)}{D}},
\end{eqnarray}
where $Z_{st}$ is a normalization constant.

\begin{figure}[t]
 \centering
   \includegraphics[width=\linewidth]{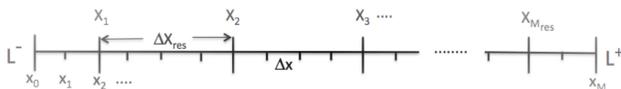}
  \caption{Scheme of the interpolation points $x_i$ and the supporting points $X_j$ used in the numerical calculations. In this scheme we have used $\lceil \frac{M}{M_{res}} \rceil=4$.}
 \label{fig:scheme}
 \end{figure}

We are interested in finding numerically the system's SPDF, $p_{st}(X_j)$, at a set of $M_{res}$ evenly spaced supporting points $X_j$ in a finite part of the physical space, given by $x \in [L^{-},L^{+}[$.  
The
region of interest is divided into $M \geq M_{res}$ subregions of size
$\Delta x=\frac{L^{+}-L^{-}}{M}$, the $i$-th subregion is bounded by $(x_i,x_{i+1})$, $i=0,\dots,M-1$, with $x_i=i\Delta x+L^{-}$. Supporting points are
given explicitly by $X_j=L^{-}+(j-\frac{1}{2}) \Delta
X_{res}$, $j=1,2,\dots,M_{res}$, with $\Delta X_{res}=\lceil \frac{M}{M_{res}} \rceil \Delta x$, see Fig. \ref{fig:scheme}. 
Here $\lceil z \rceil$ denotes the largest integer smaller than or equal to $z$.

Let us introduce the probability $P_i(t)$ for finding a particle in the
$i$-th subregion at time $t$, and the corresponding set
$\textbf{P}(t)=(P_0(t),P_1(t),...,P_{M-1}(t))$.  Initially, no
information on the system is available, thus, each subregion is given
an arbitrary amount of probability $P_i(0)$, simply fulfilling the
normalization condition $\sum_{i=0}^{M-1} P_i(0)=1$. Naturally,
equilibration can be accelerated if one already has information on
the SPDF of the system (Eq. (\ref{equ1DstochSystem})). In that case,
one can choose $\textbf{P}(0)$ close to the set $\textbf P_{st}$, which
optimally approximates the SPDF:
\begin{eqnarray}
\label{equ1DstationaryProb}
P_{st,i}=\int_{x_i}^{x_{i+1}}dx  \ p_{st}(x).
\end{eqnarray}
However, in general no such information is required.
\subsection{Time evolution} 
After setting the initial set $\textbf{P}(t=0)$, the time evolution of
the $\textbf{P}(t) \rightarrow \textbf{P}(t+h)$ is performed using
three different steps.

We start with a \textit{redistribution step}, in which $N$ walkers
(copies of the system) are uniformly distributed in each subregion.
Besides their individual positions $x^{k}_{i}(t)$, where $i=0,...,M-1$
denotes the particular subregion and $k=1,...,N$ the individual
walkers, each walker possesses a given amount of weight
$q^{k}_i(t)$. This is nothing but the present probability in the $i$-th
  subregion distributed to the $N$ walkers, which yields
\begin{eqnarray}
\label{equWeight}
q^{k}_i(t)=\frac{P_i(t)}{N}. 
\end{eqnarray}
Note that one does not need to introduce walkers in subregions with
$P_i(t)=0$.

After the redistribution step has been performed,
Eq. (\ref{equ1DstochSystem}) is integrated for all walkers, using a
Brownian dynamic simulation step $h$ and an arbitrary integration
scheme. This \textit{integration step} realizes the time evolution
$x^{k}_i(t) \rightarrow x^{k}_i(t+h)$. Here walkers transport
probability between the subregions. As walkers are independent of each other, it is of importance to note that the particular time evolution of each one of the $N\times M$ walkers is due to different sample paths in the stochastic parts of the
  Langevin equation.

Lastly, an \textit{updating step} is performed, in which the new
probabilities $P_i(t) \rightarrow P_i(t+h)$ are calculated by summing
up the weights of all walkers that are currently located in the
particular subregion,
\begin{equation}
P_i(t+h)=\sum_{i',k|x_{i'}^k(t+h)\in(x_{i},x_{i+1})}q^k_{i'}(t).
\end{equation}

In what follows, we will name the sequence of redistribution,
integration and updating step as \textit{running step}. After an
equilibration time $T_{therm}$, the set $\textbf{P}(t)$ reaches a
stationary regime, where the $P_i(t)$'s fluctuate around their mean
values $\langle P_i\rangle$.
\subsection{Calculating the stationary probability} 
The individual $\langle P_i\rangle$ are estimated by averaging over a total amount
of $N_T$ sets $\textbf{P}(t_\ell)$, $\ell=1,2,...,N_T$, taken, after the
system has reached the stationary regime, every
$n_{av}$ running steps: $t_\ell=T_{therm}+(\ell-1) n_{av} h$.  It turns
out that the mean probabilities $\langle P_i\rangle$ coincide with the stationary
probability $P_{st,i}$ (see Eq. (\ref{equ1DstationaryProb})), for
compatibly chosen time step $h$ and the size of a subregion $\Delta x$
(see Sec. \ref{convergenceCriteria}).

Finally the SPDF on the supporting points $p_{st}(X_j)$ is calculated
by adding the adjacent $\langle P_i\rangle$ and dividing by the size $\Delta X_{res}$ (in order to have a properly normalized PDF):
\begin{eqnarray}
\label{equ:EvaluSPDF}
 p_{st}(X_j)=\frac{1}{\Delta X_{res}}\sum_{i=(j-1)\lceil \frac{M}{M_{res}} \rceil}^{j\lceil \frac{M}{M_{res}} \rceil-1} \langle P_i\rangle.
\end{eqnarray}
\subsection{Calculation of the probability current} 
The stationary probability current $J_{st}(x)$ at position $x$ can be
easily calculated by adding up (with the right sign) the weights of all walkers, passing
$x$ to the right and to the left per unit time.  In practice $x$
should be the boundary of a subregion. If $x=x_{i}$
the current $J(x_{i})(t)$ is given by:
\begin{eqnarray}
\label{equ1DCurrent}
J(x_{i})(t)=\frac{1}{h} \left( \sum_{i',k \in R_i} q^{k}_{i'}(t)\,
- \sum_{i',k \in L_i}q^{k}_{i'}(t) \,\right)
\end{eqnarray}
and $R_i$ indicate these walkers which cross the boundary moving
rightwards, i.e.  $ x^{k}_{i'}(t) > x_{i} \wedge \ x^{k}_{i'}(t-h) < x_{i}$. Alternatively, $L_i$ assign walkers transporting weight
leftwards, $x_{i'}^k(t) < x_{i} \wedge x_{i'}^k(t-h) > x_{i}$.

Averaging over $N_T$ such estimates, taken in the stationary regime,
leads to the average current $\langle J(x_{i})\rangle$, which converges towards
$J_{st}(x_i)$ for $N_T \rightarrow \infty$.
\subsection{Implementation of boundary conditions} 
The implementation of boundary conditions for the probability current or the SPDF is straightforward.
Right now, absorbing boundaries are already implemented at $L^{-}$ and $L^{+}$, since walkers that pass these boundaries are
not located in any subregion. Therefore, their weights will get lost in the next updating step. Reflecting boundary conditions at $L^{+}$ can be implemented by setting $x^{k}_i(t+h) \rightarrow  2 L^{+}- x^{k}_i(t+h)$ for all walkers with $x^{k}_i(t+h) > L^{+}$. Hence, the
probability current at $L^{+}$ will be zero. Reflecting boundaries at $L^{-}$ can be implemented analogously.
\subsection{Convergence criteria}\label{convergenceCriteria} 
\subsubsection{One-dimensional systems} 
In order to ensure, that $\langle P_i\rangle$ and $\langle J(x_{i})\rangle$ converge towards the stationary probability distribution and the probability current of Eq. (\ref{equ1DstochSystem}) for $N_T \rightarrow \infty$, the time step $h$ and the size of a subregion $\Delta x$ have to 
fulfill specific criteria. This is due to the redistribution step, where walkers are uniformly distributed in each subregion.
This implicates statistical errors, since they can reach positions in a subregion that, are inaccessible or at least more improbable. Therefore, walkers can more easily escape from potential minimums or reach regions of low probability. 
This effectively flats the probability distribution, leading to more probability in regions of low probability, for instance,  around local maximums of $U(x)$, and less probability in the potentials minimums. 
In order to overcome this problem, earlier works \cite{huber1996weighted,bhatt2010steady} have stored the positions and weights of
all walkers. In the next redistribution step, walkers were only spaced on the stored positions according to the weights belonging to them. This requires a lot of computer memory, especially for large $N$ and $M$. 

However, we found that one does not need to store these information, if the subregions are small enough to ensure that walkers have a non-negligible probability to leave them during one integration step.
As a measure of how far a walker can step, due to the fluctuations, in one time step, we use the diffusion length $L_{dif}=2 \sqrt{D h}$.
Thus, the size of a subregion $\Delta x$ should be small compared to the diffusion length $L_{dif}$
\begin{eqnarray}
\label{equ1DCritBox}
\Delta x \ll 2 \sqrt{D h}.
\end{eqnarray}
The distance a walker can pass during an integration step is not only
determined by $L_{dif}$, but also by the deterministic
dynamics, leading to a step length $L_{det}=f(x) h$ in first
order.  Usually $f(x)$ changes very fast at the boundaries of
the simulation area, which produces high deterministic
velocities and regions of low probability. Walkers can only
reach these regions, if the fluctuation are strong enough to balance
the deterministic force, i.e. if
\begin{eqnarray}
\label{equ1DNoiseCompensatesGradient}
|f(x)| h < 2 \sqrt{D h}
\end{eqnarray}
for all $x \in [L^{-},L^{+}[$. This leads to a condition for the time step $h$:
\begin{eqnarray}
\label{equ1DCritTime}
h < h_{\max} := \frac{4 D}{\max_{x \in [L^{-},L^{+}]}f^2(x)}.
\end{eqnarray}
Hence, lower time steps allow one to sample regions, far from the potential extrema and for instance, the tails of the SPDF.
If a larger time step is chosen, the $\langle P_i\rangle $ will run to zero in 
subregions with larger deterministic force. 

Since it is often difficult to fulfill Eq. (\ref{equ1DCritTime}) in the entire simulation area, one should choose a time step, which allows one to fulfill Eq. (\ref{equ1DCritBox}).
\subsubsection{Multidimensional systems} 
In general, our method can be applied to stochastic dynamical systems
in arbitrary dimension $d \geq 1$
(Eq. (\ref{equ1DstochSystem})). For the foundation of the
  algorithm, we refer to the Appendix \secref{sec:found}.  However,
in order to ensure convergence in a finite region $A$, the criteria
for the time step $h$ and the size of a subregion $\Delta x_i$ along
the $i$-th direction should be fulfilled.  For noise dominated
directions the criteria hold, therefore, the condition for the time
step (Eq. (\ref{equ1DCritTime})) becomes:
\begin{eqnarray}
\label{equmultiDCritTime}
h \approx \min_{1\le n\leq d}(\frac{4 D_n}{\max_{A}(f_n(\textbf x))^2}),
\end{eqnarray}
and the criteria for the size of a subregion along the $n$-th directions (Eq. (\ref{equ1DCritBox})) reads:
\begin{eqnarray}
 \Delta x_n \ll 2 \sqrt{D_n h}.
\end{eqnarray} 
However, there might be directions without any noise ($D_n=0$). 
In that case the length of a subregion should ensure, that walkers can leave it due to the deterministic term $f_n(\textbf x)$, leading to 
\begin{eqnarray}
\label{critnoNoise}
\Delta x_n \leq f_n(\textbf x) h.
\end{eqnarray} 
otherwise information on the deterministic dynamics gets lost during the redistribution step, since walker that stay into a subregion do not produce any change in the $P_i$ and are again randomly placed in their subregion during the next redistribution step.
For equally sized subregions $\Delta x_n$ should be the minimum value of Eq. (\ref{critnoNoise}) with respect to all $\textbf x$ in the simulation area for each of the $d$ directions. 
These criteria can lead to a huge number of subregions, especially in high dimensional spaces.
 
To appropriately reduce the number of subregions, the $\Delta x_n$ should be chosen in order to locally fulfill the criteria. This was implemented by a \textit{grouping algorithm}, that groups original subregions into "larger" ones as long as 
walkers can leave these due to the deterministic term (Eq. \ref{critnoNoise}). Depending on the system, this procedure highly reduces the total number of "larger" subregions $M_{group}$. 
If the \textit{grouping algorithm} was used, $N$ walkers are randomly placed in each of these 
"larger" subregions and a probability $P_i$ of finding a walker in the corresponding area was introduced. We find that such grouping highly reduces the computational costs, since less subregions and therefore less walkers are required. 
Since walkers jump out of these regions until the next redistribution step starts, the algorithm still approximates the correct SPDF. 
\subsection{Simulation techniques} 
Simulations were performed on a Intel\textregistered Xeon
\textregistered CPU E31245 @ 3.30GHz processor with 16 Gb DDR-3
RAM. The algorithm described above was implemented in a $C^{++}$
program for one- and two-dimensional systems.  Runs of the algorithm
are specified by the time step $h$, the size of a subregion $\Delta
x$, ($\Delta y$, in two-dimensional problems), the simulation area, given by $L^{-}$ and $L^{+}$
($L^{\pm}_x$, $L^{\pm}_y$), the number of walkers per subregion $N$,
the thermalization time $T_{therm}$, and the number of running steps
between two sets of $\textbf{P}$ denoted as $n_{av}$. The numerical
integration of the Langevin Eq. (\ref{equ1DstochSystem}) was done
using a Heun scheme.  A resolution of $M_{res}=200$ was used in any
direction.

To compare the results with other methods, we also perform Brownian
dynamics simulation (BDS) using $N_{Brown}$ initially uniformly placed
particles in the simulation area. Integration was done using the Heun
scheme \cite{kloeden2011numerical,gilsing2007sdelab} with integration time step
$h_{Brown}$.  After a thermalization time $T_{therm,Brown}$ the
particles positions were recorded after time intervals $\Delta
t_{Brown}$. The BDS was given a running time $T_{run}$ (real CPU time)
which usually equals the time our algorithm needs to produce its
results. After $T_{run}$ the BDS was stopped and the SPDF was
calculated using the recorded particle positions.  We set
$T_{therm,Brown}=T_{therm}$, $h_{Brown}=h$, $\Delta t_{Brown}=n_{av}
h$ to make results comparable.
\section{Model systems and results} 
\subsection{One dimensional system}\label{1Dsystem}
In order to demonstrate the implementation of the algorithm, we
  study overdamped Brownian motion in a bistable potential
  $U(x)=-\frac{x^2}{2}+\frac{x^4}{4}$. Correspondingly, we put
  $f(x)=x-x^3$ and $g(x)=\sqrt{2D}$ in Eq. (\ref{equ1DstochSystem})
  which results in a bistable system which is often used to study
  bistable systems or stochastic resonance therein
  \cite{risken,gammaitoni1998stochastic}. The two stable states come
up to the potentials minimums, located at $x=-1$ and $x=1$,
respectively.  The corresponding SPDF is given by
Eq. (\ref{equ:bistablePDF}).  For low noise strength, the SPDF attains
sharp peaks at the potentials minimums and decreases down to low
values at the borders and the local maximum, for instance for $D=0.01$
$p_{st}(0)\approx 10^{-11}$.
\subsubsection{Equilibration} 
\begin{figure}[t]
 \centering
   \includegraphics[width=\linewidth]{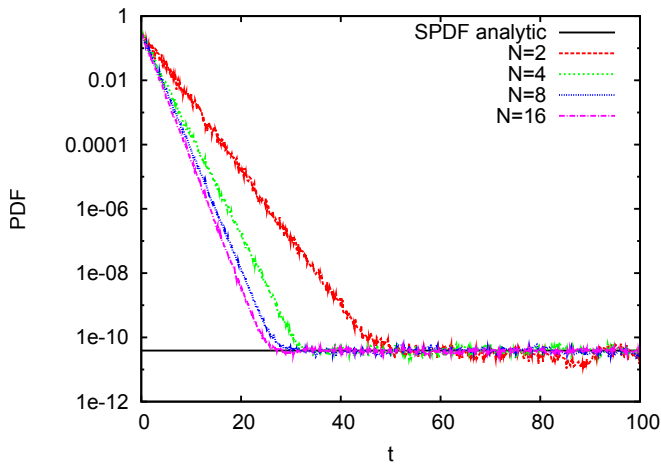}
  \caption{$PDF$ at $x=0$ obtained from $P_i$ by $PDF(t)=\frac{P_i(t)}{\Delta x}$ for the subregion containing $x=0$ for $D=0.01$ and different numbers of walkers per subregion $N$. Parameters are chosen as in run 1 (see Tab. \ref{tabValuesCriteriaBistable}).}
 \label{fig:Fluctuations}
 \end{figure}

At first, we study the equilibration process, performing simulations
with different numbers of walkers per subregion $N$.  Results are shown in
Fig. \ref{fig:Fluctuations}. Analyzing the time dependence of the
probability $P_i(t)$, we find that longest thermalization time occurs at the local
maximum of $U(x)$.
Note that runs with larger $N$ thermalize at lower $t$, but one needs more integration steps. 

After thermalization has been achieved, we evaluate the coefficient of variation of the probability, given by 
\begin{align}
\label{equ:coefficientOfVary}
c(x_i)=\frac{\sqrt{\langle (P_i-\langle P_i\rangle)^2\rangle}}{\langle P_i\rangle}
\end{align}
where averages $\langle \cdots\rangle$ are performed for a fixed number $N_T$ of sets ${\bf P}$ and different $N$ and $x_i$. We find 
it to scale according to $\frac{1}{\sqrt{N}}$ (data not shown). 
\subsubsection{Stationary probability density function} 
We start to calculate the SPDF $p_{st}$ in the region $[L^{-},L^{+}[$ for a small noise strength ($D=0.01$). 
The time step $h$ and the box size $\Delta x$ are set according to the criteria (see Tab. \ref{tabValuesCriteriaBistable}).
\begin{table}
\begin{center}
 \begin{tabular}[t]{|l|c|c|c|c|l|}
\hline
\hline
    run & $L^{-}$ & $L^{+}$ & $h$ & $\Delta x$ & $M$\\
\hline
\hline
    1 & -1.4 & 1.4 & 0.011 & 0.00104869 & 2670 \\
\hline
    2 & -1.75 & 1.75 & 0.0015 & 0.000387297 & 9037\\
\hline
    3 & -2.5 & 2.5 & 0.0001 & 0.00010775 & 46404\\
\hline
\hline
  \end{tabular}
\end{center}
\caption{Time step $h$ and box size $\Delta x$ according to convergence criteria Eq. (\ref{equ1DCritTime}) and (\ref{equ1DCritBox}), where we choose \mbox{$h=\frac{h_{max}}{2}$} and \mbox{$\Delta x = \frac{1}{20} L_{dif}$}.}
\label{tabValuesCriteriaBistable}
\end{table}
Using the results shown in \mbox{Fig. \ref{fig:Fluctuations}}, we set the thermalization time $T_{therm}=50$ for a run with $N=2$. Time averages after thermalization were calculated over an ensemble of $N_T = 10^4$ sets $\textbf{P}$.
\begin{figure}
  \centering
    \includegraphics[width=\linewidth]{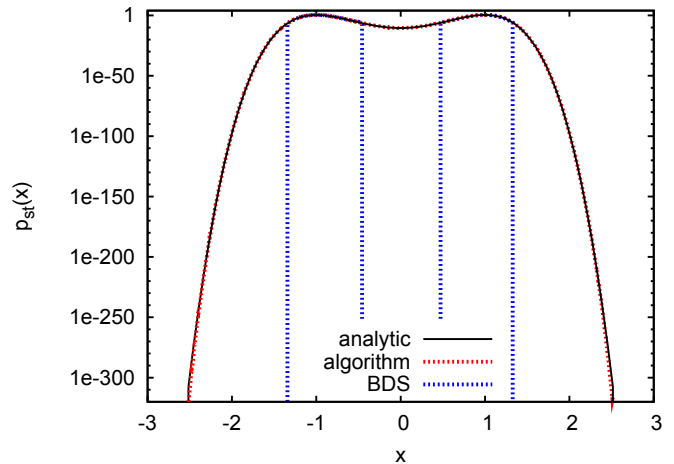}
  \caption{Estimates for the stationary probability density obtained from the algorithm for run $3$ (see Tab. \ref{tabValuesCriteriaBistable}) and by using a Brownian dynamics simulation for $D=0.01$. 
Analytic results are obtained from Eq. (\ref{equ:bistablePDF}).}
  \label{fig:bistablePDF}
\end{figure}
Results for the SPDF are shown in Fig. \ref{fig:bistablePDF}. The algorithm calculates the tails of the distribution down to $10^{-300}$ correctly, after a running time $T_{run} \approx 27 \ hours$. We also calculate the SPDF using Brownian dynamics simulation using $N_{Brown}=10^4$, which stops estimating at a level of  $10^{-6}$ after the same running time. 
Further runs were performed (see Tab. \ref{tabValuesCriteriaBistable} (run $2$) and (run $3$)), approximating the tails down to $10^{-10}$ ($M_{group}=1136$) and $10^{-48}$ ($M_{group}=6789$) after a running time of $\approx 30 \ sec$  and $\approx 10 \ min$ , 
respectively (data not shown).

Simulations for different values of $\frac{\Delta x}{L_{dif}}$ indicate that insignificant deviations from the analytic SPDF occur for $\Delta x > \frac{1}{20}$.
\subsubsection{Probability current} 
Next, we present that our algorithm can be used to calculate the
escape rate to pass the energy barrier at $x_{max}=0$.  Such problems
are typical for chemical reactions \cite{hanggi_rmp90} and in the field of neuroscience
\cite{tuckwell}.

Initially, only $N$ particles are assigned at the subregion including $x_{min}=1$, so approximating an initial delta-like probability distribution for $t=0$. Furthermore, an absorbing boundary right behind the local maximum ($x_{abs}=-0.01$) is included. To fulfill normalization of the SPDF, walkers that reach $x_{abs}$ are reinjected immediately at $x_{min}$.
The escape rate to pass the barrier is given by the probability current $J(x_{max})$.
For small noise intensities, the probability current on top of the potential barrier can be described using Arrhenius law, namely:
\begin{eqnarray}
 J(x_{max})\propto e^{-\frac{\Delta U}{D}},
\end{eqnarray}
where $\Delta U=U(x_{max})-U(x_{min})=0.25$.
Since strong fluctuations are rare, but possible, we will use $J(x_{abs})$ to approximate $J(x_{max})$.
Probability currents $J(x_{abs})$ were recorded for each time step and averaged over a sequence of $n_{av}=\lceil \frac{0.1}{h} \rceil$ running steps, resulting in $\langle J(x_{abs})\rangle $.
\begin{figure}[t]
  \centering
    \includegraphics[width=\linewidth]{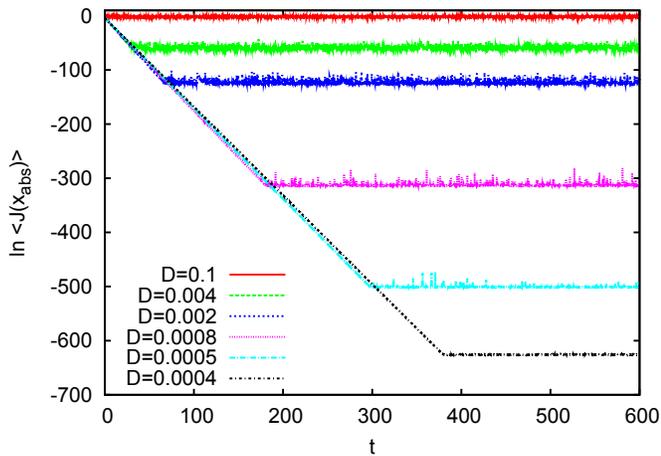}
  \caption{Time dependence of the probability current $J(x_{abs})$ obtained from our algorithm for $L^{-}=-0.02$, $L^{+}=1.39$ and decreasing noise intensities (from top to bottom). 
The time step is set $h=\frac{h_{max}}{2}$ and $\Delta x= \frac{1}{20} L_{dif}$. 
Note that $h_{max}$ and $L_{dif}$ vary according to Eq. (\ref{equ1DCritTime}) and (\ref{equ1DCritBox}), respectively, resulting in larger running times for smaller $D$.}
  \label{fig:equbistableCurrent}
\end{figure}
Figure \ref{fig:equbistableCurrent} shows the time dependence of $\ln \langle J(x_{abs})\rangle $. After a relaxation regime, where the current decays exponentially, the current reaches its stationary value.
\begin{figure}[t]
  \centering
    \includegraphics[width=\linewidth]{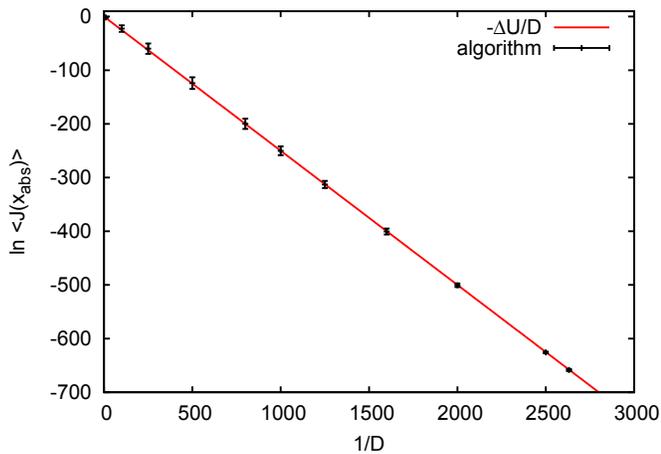}
  \caption{Stationary probability current as a function of the inverse noise strength obtained by time averaging the data partly shown in Fig. \ref{fig:equbistableCurrent} in the stationary regime.
Error bars show three standard deviations of the stationary data.}
  \label{fig:bistableCurrent}
\end{figure}
The values of \mbox{$\ln \langle J(x_{abs})\rangle $}, averaged over the stationary regime, are shown in Fig. \ref{fig:bistableCurrent} for different noise intensities. 
Fulfilling the criteria described above, the algorithm reproduces well Arrhenius law down to $\ln\langle J(x_{max})\rangle \approx -650$ corresponding to a current $J(x_{max})\approx 10^{-286}$.
\subsubsection{Efficiency compared to weighted-ensemble Brownian dynamics simulation} 
In order to compare the efficiency of two algorithms, important quantities are the transient time required to first reach the steady state. Once the algorithm reaches the steady state,
we quantify the size of the fluctuations by the coefficient of variation Eq. (\ref{equ:coefficientOfVary}) at the potential's local maximum $x=0$. The computation time mainly depends on the
number of integrations $N_{int}$ needed, to reach the stationary regime.
\begin{figure}[t]
  \centering
    \includegraphics[width=\linewidth]{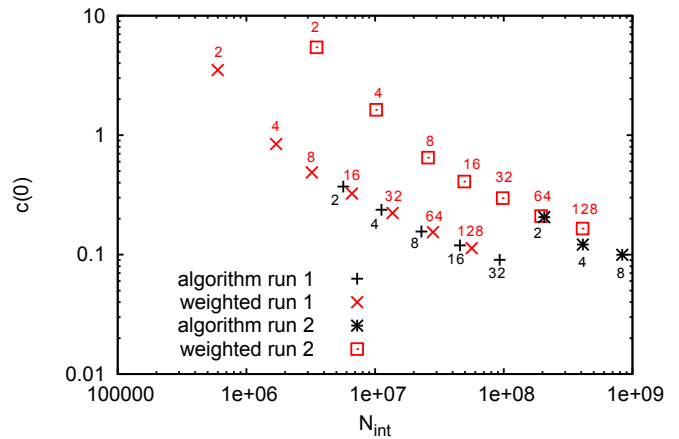}
  \caption{Relative fluctuations plotted over the number of integration steps, needed for equilibration, for the algorithm (black) and standard (WE) simulation (red). The number of walkers per subregion is shown for each point.
Simulations were done for run $1$ and $2$ (compare Tab. \ref{tabValuesCriteriaBistable}).}
  \label{fig:Efficiency}
\end{figure}
In order to compare the efficiency, the size of fluctuations (Eq. (\ref{equ:coefficientOfVary})) in the local minimum relative to $p_{st}(0) \approx 3.88717\times 10^{-11}$ during the stationary regime is plotted over $N_{int}$ in Fig. \ref{fig:Efficiency}. 
The most effective algorithm would be located close to the origin. Comparing the efficiency of standard WE simulations and our algorithm, we find that WE simulations with low $N$ equilibrate faster. However,
the precision highly depends on the fluctuations during the stationary regime. To produce results of same precision (same $c(0)$) both algorithms approximately need the same $N_{int}$.
Usually WE simulations were performed using thousands of walkers per subregions, resulting in a low value of $c(0)$. Here runs of our algorithm producing the same $c(0)$ need much less $N$ and have 
memory requirements independent of $N$.
\subsection{Two-dimensional systems} 
\subsubsection{Poincar\'e Oscillator} 
As an example of a two-dimensional system with known SPDF, we consider
the Poincar\'e oscillator \cite{ebeling1980influence,EbHe86,LekLsg88},
represented by the dynamical system:
\begin{eqnarray}
\begin{array}{cl}
\dot{x}= &  y\\
 \dot{y}= & (\alpha-x^2-y^2)y-x+\sqrt{2 D} \xi(t),\\
\end{array}
\label{equ:Poincare}
\end{eqnarray}  
where $\xi(t)$ represent delta correlated white Gaussian noise with zero mean. 
Using the energy function $H(x,y)=\frac{1}{2}(x^2+y^2)$, which only depends on the distance to the origin, one can calculate the associated SPDF:
\begin{eqnarray}
\label{equ:SPDFVanDerPol}
 p_{st}(x,y)=Z^{-1}_{st} \,\exp\Large(\frac{\alpha H(x,y)-H^2(x,y)}{D}\Large),
\end{eqnarray} 
(see appendix \secref{SPDFVanderPol}).
Since noise only applies to the $y$-direction, the lengths of a subregion $\Delta x$ and $\Delta y$ in $x$- and $y$-direction are calculated by Eq. (\ref{critnoNoise}) and (\ref{equ1DCritBox}), respectively.
The minimum of $|f_{x}(x,y)|=|y|$ is equal to $0$, therefore, we choose $\Delta x = \Delta y \ h$, which corresponds to a first order approximation of $|f_{x}(x,y)|$ in the next subregion. 
\begin{figure}[t]
  \centering
    \includegraphics[width=\linewidth]{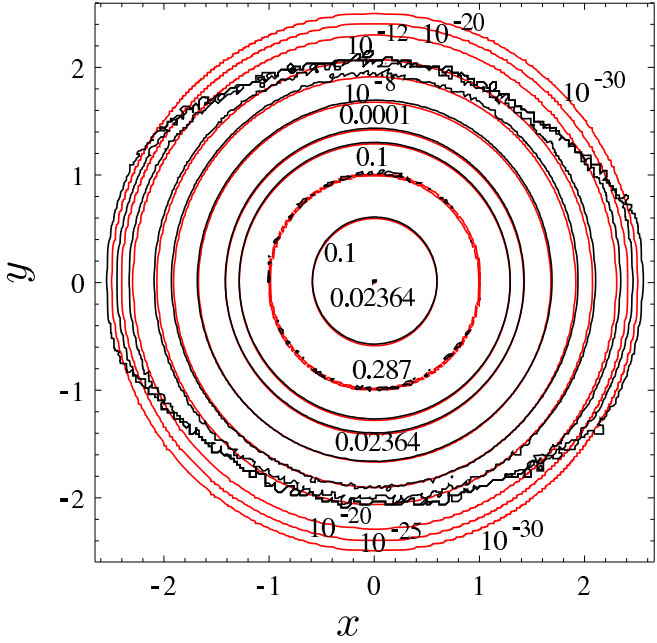}
  \caption{Contour plots of the SPDF ($D=0.1$ and $\alpha=1$) obtained from the algorithm (black), using the parameters 
$L_x^{-}=L_y^{-}=-3$, $L_x^{+}=L_y^{+}=3$, $T_{therm}=8$, $N_T=10^3$, $h=0.01$, $M_x=189737$, $M_y=1897$, $M_{group}=1688940$, \mbox{$N=2$}, ($\Delta y=\frac{1}{20} L_{dif}$) and the analytic solution (red) \mbox{Eq. (\ref{equ:SPDFVanDerPol})}. 
Contour lines are labeled according to represented values of the SPDF and show the rotational symmetry. The SPDF possesses its global maximum at $H(x,y)=\frac{\alpha}{2}$, corresponding to the unit circle for our choice of $\alpha$, and a local minimum in the origin.  Running time $\approx 5 \ hours$.}
  \label{fig:VanPol_Contour}
\end{figure}
Results for the SPDF are shown in Fig. \ref{fig:VanPol_Contour}. Interestingly, the algorithm approximates  better the SPDF along the direction where no noise was applied. Here the SPDF is sampled down to $10^{-30}$. We found that the algorithm 
slightly oversamples the analytic SPDF in the tails. This is due to the statistical errors, made during the redistribution step. By reducing the size of a subregion, this error can be reduced further. 
Along the $y$-direction, noise is applied. Here the behavior is similar as in the one dimensional example (see above). For runs with larger $\Delta y$ (results not shown) the algorithm slightly oversamples the SPDF in the origin.
\subsubsection{Bistable system with colored noise} 
As a further example, we calculate the SPDF of the two dimensional system:
\begin{eqnarray}
\label{equ:bistableColor}
\begin{array}{cl}
 \dot{x}= & x-x^3+y\\
 \dot{y}= &-\displaystyle \frac{1}{\tau}y+\frac{1}{\tau}\sqrt{2 D}\xi(t),\\
\end{array}
\end{eqnarray}
where $\tau$ denotes the time scale separation between $x$ and the
colored noise $y$. The white Gaussian noise $\xi(t)$ has been already
described above.  This system has been studied previously in
\cite{debnath1990holes,HaJu95}. Like in the bistable system we have
studied above, the SPDF has maximums at $(x,y)=(1,0)$ and
$(x,y)=(-1,0)$. However, for some combinations of $\tau$ and $D$, the
SPDF possesses a local minimum at $(x,y)=(0,0)$.
\begin{figure}[t]
\begin{minipage}{\linewidth}
  \centering
    \includegraphics[width=0.8\linewidth]{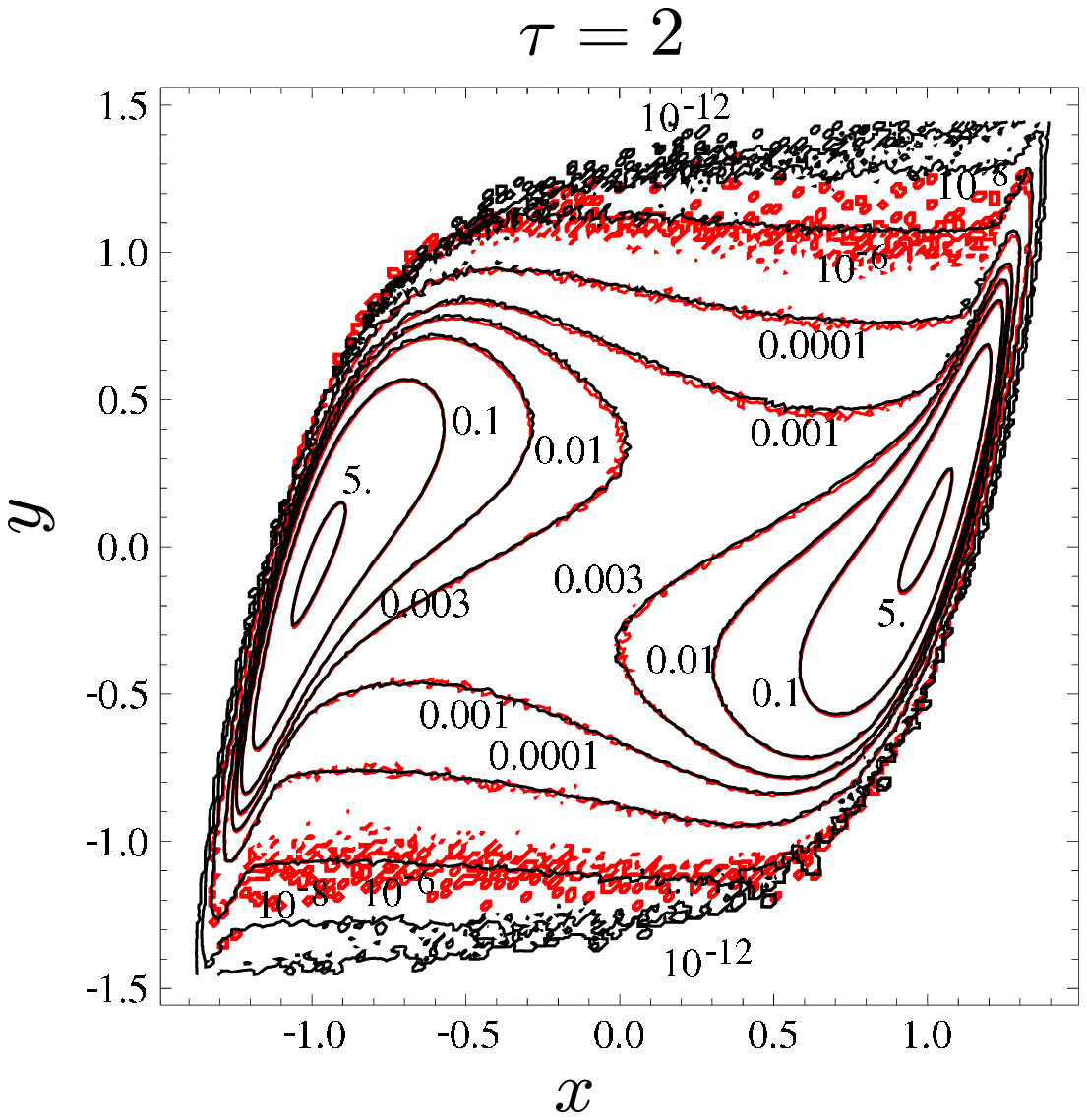}
\vspace{1cm}
\end{minipage}
\begin{minipage}{\linewidth}
  \centering
    \includegraphics[width=0.8\linewidth]{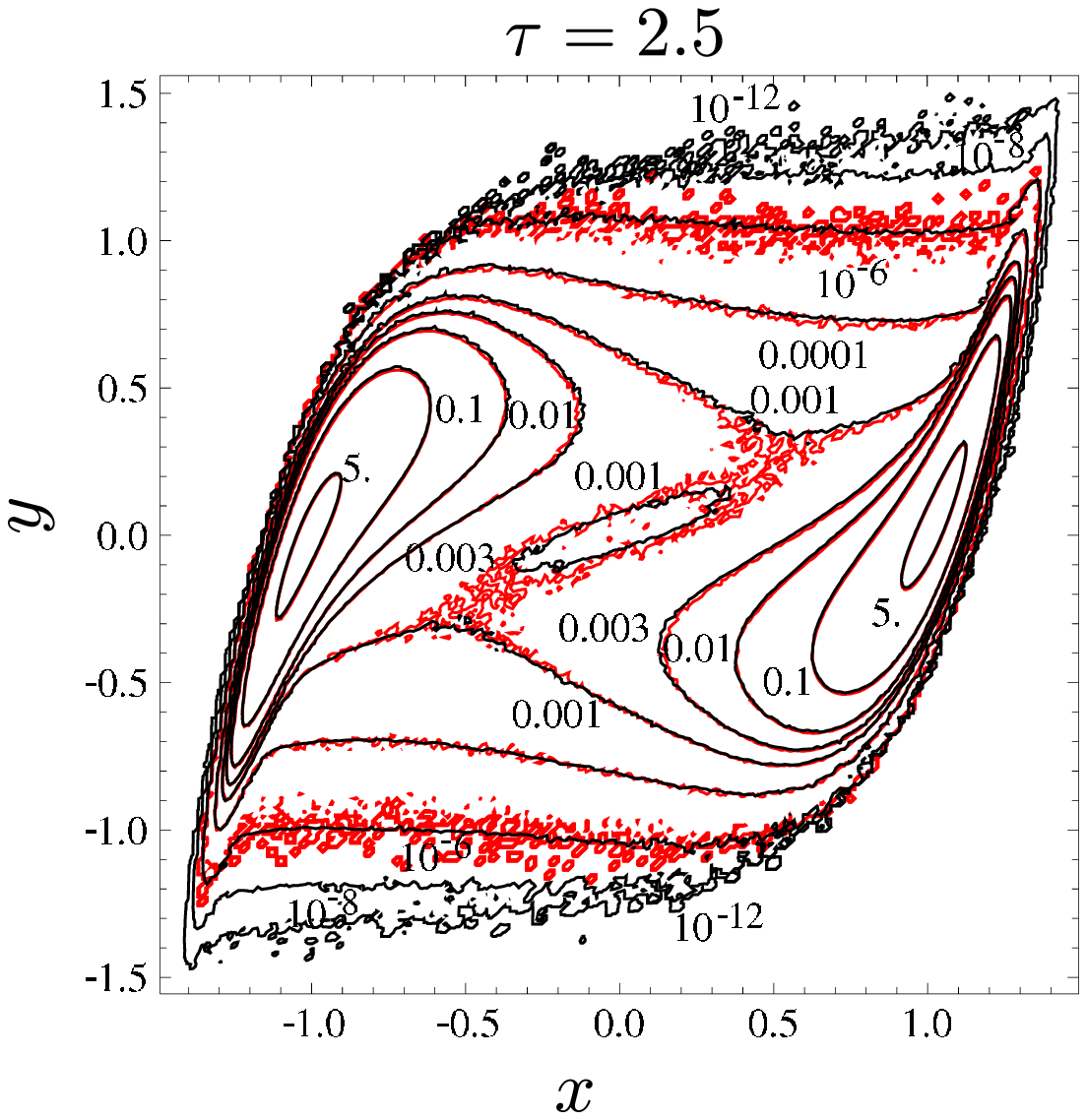}
\end{minipage}
  \caption{Contour plots of the SPDF of Eq. (\ref{equ:bistableColor}) ($D=0.1$, $\tau=2$ and $\tau=2.5$) obtained from the algorithm (black) and BDS (red) of the same running time.
Parameters: (top) $L_x^{-}=L_y^{-}=-1.5$, $L_x^{+}=L_y^{+}=1.5$, $T_{therm}=10$, $N_T=10^3$, $h=0.0889$, $M_x=14319$, $M_y=1273$, $M_{group}=165003$, $N=2$,($\Delta x = \Delta y \ h$, $\Delta y=\frac{1}{20} L_{dif}$), $N_{Brown}=10^3$, and
(bottom) $L_x^{-}=L_y^{-}=-1.5$, $L_x^{+}=L_y^{+}=1.5$, $T_{therm}=12$, $N_T=10^3$, $h=0.0889$, $M_x=17899$, $M_y=1591$, $M_{group}=206957$, $N=2$,($\Delta x = \Delta y \ h$, $\Delta y=\frac{1}{20} L_{dif}$), $N_{Brown}=10^3$. Running times are $19 \ min$ (top) and $21 \ min$ (bottom)}
  \label{fig:OUtau2and25bistableCurrent}
\end{figure}
Plots of the SPDF are depicted in Fig. \ref{fig:OUtau2and25bistableCurrent}. Our algorithm samples the SPDF down to $10^{-12}$, whereas BDS breaks down at a level of $10^{-6}$. The minimum, occurring for $\tau=2.5$ was clearly found by the algorithm.
\subsubsection{FitzHugh-Nagumo-system} 
As a last example, we consider the widely used FitzHugh-Nagumo-system \cite{fitzhugh1961impulses}, which is often used in the field of Neuroscience \cite{lindner2004effects} or to 
study synchronization \cite{gunton2003role,toral2003characterization} and coherence phenomena \cite{toral2007system}, represented by:
\begin{eqnarray}
\label{equ:FHN}
\begin{array}{cl}
 \dot{x}= & \frac{1}{\epsilon}(x-x^3-y)+\sqrt{2 D_x}\xi_x(t)\\
 \dot{y}= & \gamma x-y+b+\sqrt{2 D_y}\xi_y(t).\\
\end{array}
\end{eqnarray} 
Here $\epsilon$ denotes the timescale separation between the activator variable $x$ and the inhibitor variable $y$. $\xi_x(t),\xi_y(t)$, represent independent zero-mean delta-correlated Gaussian white noises. We want to study the stationary probability density in the case of $D_x=D_y=D$ for a time scale separation $\epsilon=0.1$.
We set the parameters according to Ref. \cite{kostur2003stationary} to $b=1.4$, $\epsilon=0.1$ and $\gamma=2$. Thus, the system is in the excitable regime. 
Since the deterministic part of the equation for the activator variable increases very fast if $x$ is increased, we have to choose a time step $h=0.01$, which is small enough, that the walkers' steps are small compared to $1$, but 
allows us to fulfill the criteria for the size of the subregions. 
\begin{figure}[t]
\begin{minipage}{\linewidth}
  \centering
    \includegraphics[width=0.8\linewidth]{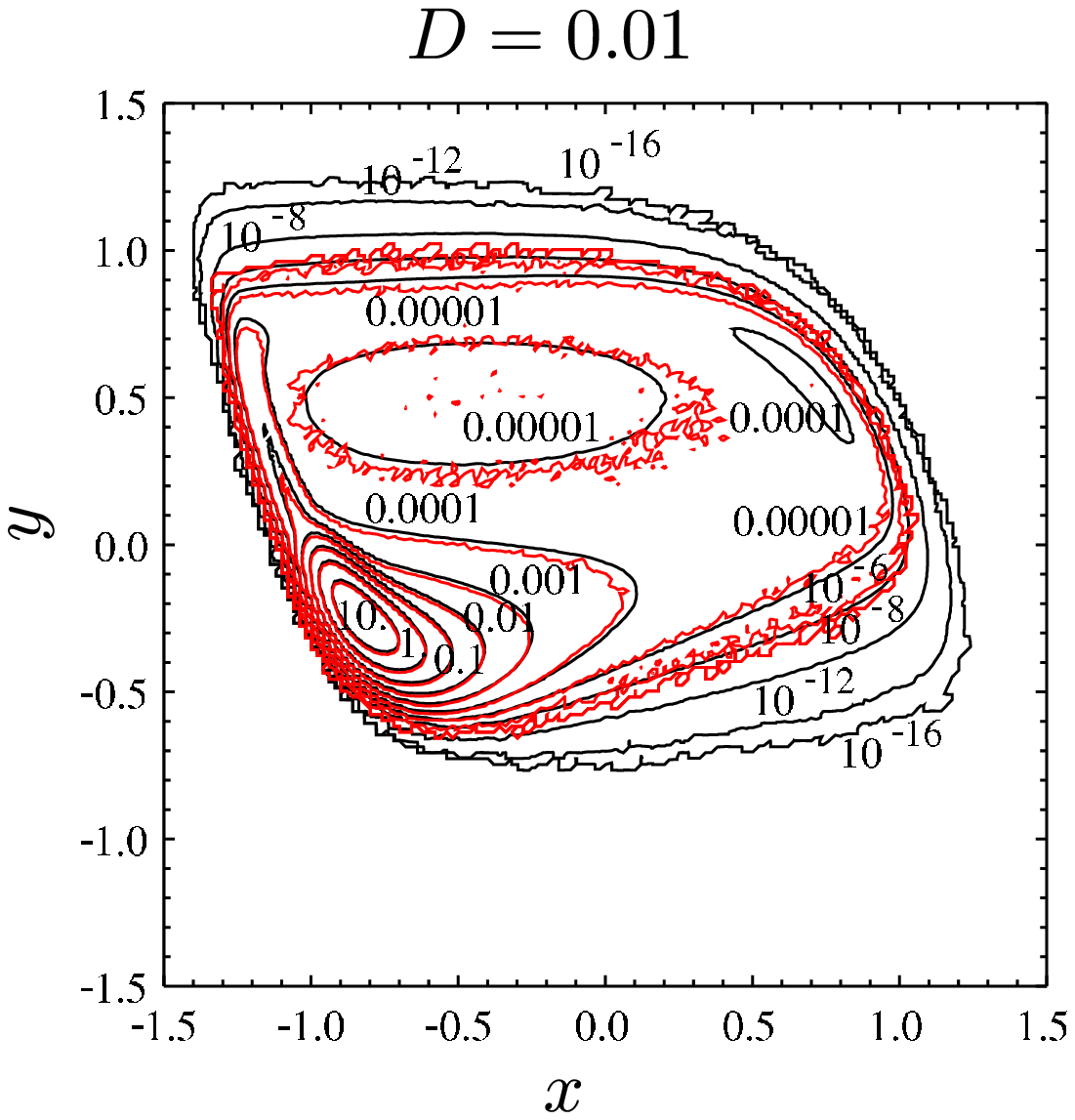}
\end{minipage}
\begin{minipage}{\linewidth}
   \centering
    \includegraphics[width=0.8\linewidth]{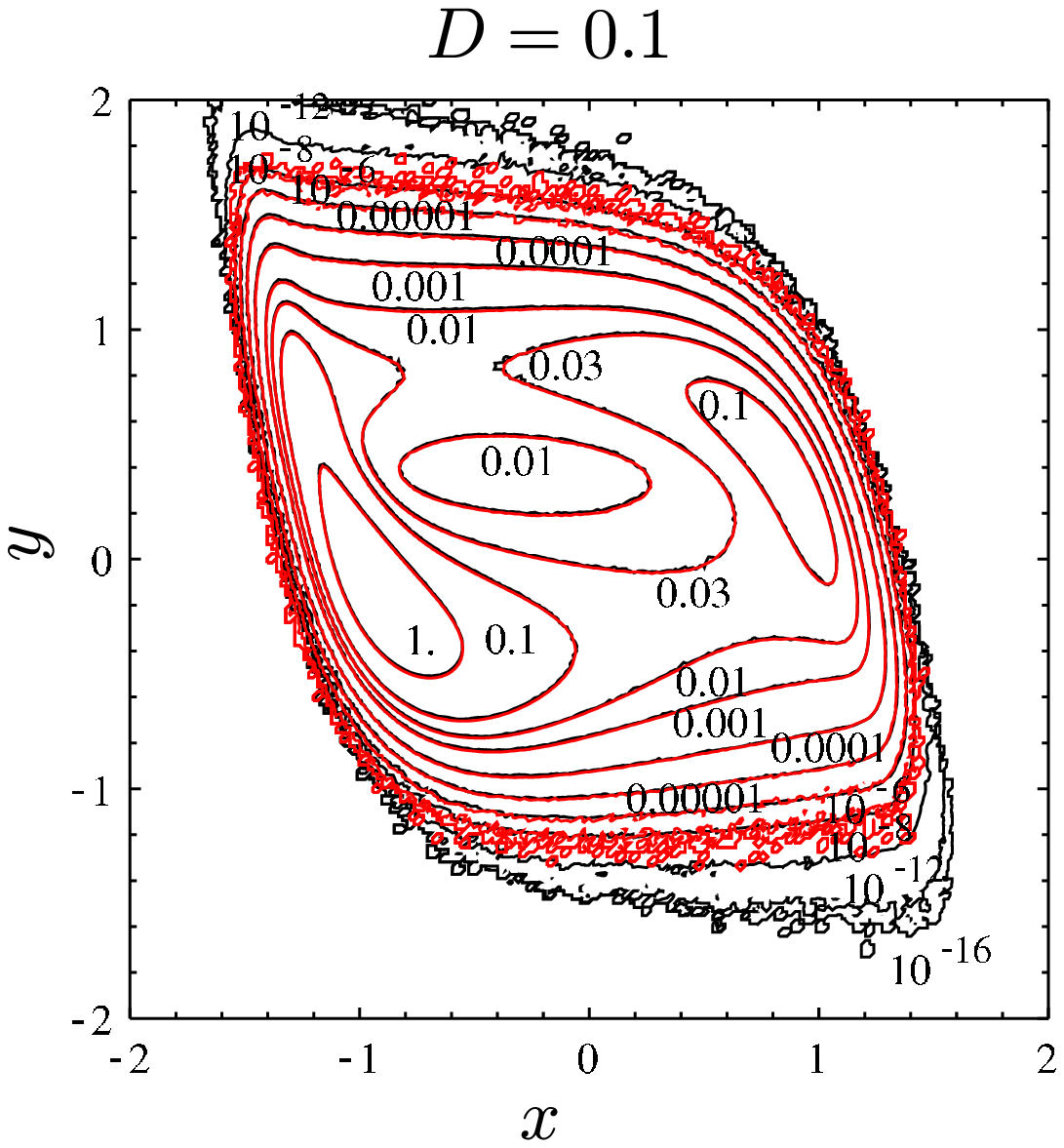}
\end{minipage}
  \caption{Contour plots of the the SPDF of Eq. (\ref{equ:FHN}) fod ($D=0.01$ (top) and $D=0.1$ (bottom)) obtained from the algorithm (black) and BDS (red) with the same running time.
 Parameters: (top) $L_x^{-}=L_y^{-}=-2$, $L_x^{+}=L_y^{+}=2$, $T_{therm}=5$, $N_T=10^3$, $h=0.01$, $M_x=4000$, $M_y=4000$, $N=2$,($\frac{\Delta x}{\sqrt{D h}}=\frac{1}{10}$), $N_{Brown}=10^3$, $M_{group}=314027$, and 
(bottom) $L_x^{-}=L_y^{-}=-2$, $L_x^{+}=L_y^{+}=2$, $T_{therm}=3$, $N_T=10^3$, $h=0.01$, $M_x=1333$, $M_y=1333$, $N=2$, ($\Delta x= \Delta y=\frac{1}{20} L_{dif}$), $N_{Brown}=10^3$, $M_{group}=146568$. Running times are $27 \ min$ (top) and $19 \ min$ (bottom). 
Simulation times for
the BDS were chosen three times larger.}
  \label{fig:FHNeps01}
\end{figure}
Contour plots of the SPDF are shown in Fig. \ref{fig:FHNeps01}. Especially regions of low probability are much better sampled, using the algorithm.
In the case of low diffusion ($D=0.01$, Fig. \ref{fig:FHNeps01} (top)) the algorithm runs down to $10^{-16}$,  whereas BDS stops at a level of $10^{-6}$. Especially the minimum
is much better sampled by our algorithm. Note that the local maximum located in the surroundings of $(x,y)=(0.7,0.5)$ was not found by BDS. For higher diffusion values ($D=0.1$, Fig. \ref{fig:FHNeps01} (top)), significant differences
can be found only in the tails.
\section{Discussion}
\subsection{One dimensional system}
In the one dimensional system (see section \ref{1Dsystem}) we succeeded to approximate the SPDF down to 
$10^{-300}$. If equally sized subregions are used, a huge number of subregions has to be implemented, in order to fulfill the convergence criteria (see section \ref{convergenceCriteria}).
However, by evaluating the SPDF according to Eq. (\ref{equ:EvaluSPDF}) the additional computational costs also reduce the fluctuations of the estimated SPDF.   
If the subregions are to large, the theoretical SPDF is overestimated by the algorithm in potential minimums and underestimated in the potentials maximums. 

We also managed to calculate escape rates of size $10^{-286}$. Although the finite size of a subregion leads to small errors, in the calculated SPDF, Arrhenius law is well reproduced for such small probability currents.

At the stationary regime the $P_i$ fluctuate around their mean value, estimating the SPDF. By either increasing the number of averages $N_T$ or the number of walkers per subregion $N$ these fluctuations can be reduced, leading to higher precision. The estimation error scales with $\sqrt{N_T}$ and $\sqrt{N}$, respectively.
However, increasing the number of walkers per subregion also effects the computational costs during the thermalization. Simulations for different $N$ show that runs with higher $N$ become stationary faster,  but this does not compensate for the additional computational costs.  
We also find, that increasing $N$ slightly improves the sampling of the SPDF's tails.
Once the system is in the stationary regime the increase of the computational costs scale linearly with $N$ and $N_T$. 
An advantage of simulations with small $N$ is, that one does not need to perform running steps for subregions with $P_i=0$ and
the number of such regions naturally increases for small $N$. We usually use the smallest possible $N=2$ and scale the estimation error by increasing $N_T$.
\subsection{Comparison with weighted-ensemble Brownian dynamics} 
Since the general idea of our method was adapted from prior simulation techniques known as weighted-ensemble (WE) sampling \cite{huber1996weighted,bhatt2010steady,bhatt2012adaptive}, we 
want to discuss advantages and disadvantages of our algorithm in comparison with these techniques. The main difference in WE techniques is the redistribution step. Here, using WE techniques, positions and weights of all walkers are stored, 
and new walkers are introduced on the stored positions considering their particular weights. 

In our algorithm, there is no need to store any position or weight, since walkers are randomly placed in each subregion. The resulting statistical errors can be neglected,
if the size of the subregions fulfills conditions which, unfortunately, lead to much larger numbers of subregions. By averaging the probabilities, these extra computational costs contribute
to an reduction of fluctuations in the stationary regime.
 
The comparison of the computational costs until equilibration and the achieved precision shows, that both algorithms posses the same efficiency for high precision runs.
\subsection{Two dimensional systems}
In the case of two dimensional systems, we found that the algorithm outperforms Brownian dynamics simulations. However, the number of subregions needed according to the criteria can be really high, especially if there are some directions without any noise. 
Here it is necessary to size the subregions to fulfill the criteria locally. A first step in that direction has already been done by the implementation of the grouping algorithm. 
We are quite confident that it is possible to reduce the computational costs by optimizing the Mesh. 
The analyzed examples show, that the running times highly depend on the investigated system. 
For the bistable system with colored noise and the FitzHugh-Nagumo-system, we obtained good results within $\approx 20$ minutes, whereas the algorithm needed about $5$ hours for the Van-der-Pol oscillator.  
\section{Conclusion}
We provided and tested an algorithm that allows the calculation of low probabilities and low rates. The algorithm is based on WE Brownian dynamic simulations, but uses a uniform distribution of walkers within each subregion.
To our findings, the resulting statistical errors can be neglected if one uses subregions small compared to the diffusion length. 
In contrast to WE methods, the required memory does not depend on the number of walkers, which leads to less memory requirements for runs with large numbers of walkers.

Special attention was payed to non-equilibrium dynamical systems.
Applying the method to one- and two-dimensional model systems, we analyze its efficiency compared to standard Brownian dynamics simulation.
Our method outperforms Brownian dynamics simulation by several orders of magnitude and its efficiency is comparable to weighted-ensemble Brownian dynamic simulations in
all studied systems and lead to impressive results in regions of low probability and small rates.

\acknowledgements This paper was developed within the scope of the
IRTG 1740 / TRP 2011/50151-0, funded by the DFG / FAPESP. LSG
acknowledges IFISC of the University of Balearic Island for cordial
hospitality and support. He also acknowledges Dr. Volkhard Bucholtz
from Logos-Verlag Berlin for earlier participation and work on the
project. RT acknowledges financial support from MINECO (Spain), Comunitat Aut\`onoma de les Illes
Balears, FEDER, and the European Commission under project FIS2007-60327, as well as the warm hospitality at Humboldt University.

\appendix
\section{Stationary probability density of the Poincar\'e oscillator} 
\label{SPDFVanderPol}
Using the energy function $H(x,y)=\frac{1}{2}(x^2+y^2)$, we get from
\eq{equ:Poincare} to a representation as a canonic dissipative
system:
\begin{eqnarray}
  \begin{array}{cl}
    \dot{x}= &  \partial_y H\\
    \dot{y}= & \partial_y(\alpha H -H^2)-\partial_x H+\sqrt{2 D} \xi(t).\\
  \end{array}
\end{eqnarray}  
The corresponding Fokker Planck equation in the $x,y$ phase space for
the SPDF $p_{st}(x,y)$ reads \cite{ebeling1980influence}:
\begin{eqnarray}
\begin{array}{cl}
  \partial_t p_{st}= 0= & -\partial_y H \partial_x p_{st}+ \partial_x H \partial_y p_{st} \\
  & - \partial_y(\partial_y(\alpha H -H^2) p_{st}) + D \partial_y^2 p_{st}\,. \\
\end{array}
\end{eqnarray}
Using the ansatz $p_{st}(x,y)=p_{st}(H(x,y))$, the first two items at
the r.h.s. cancel. The remaining second line can be integrated
once. Assuming an exponentially decaying SPDF at infinitely large
energies yield the disappearance of the irreversible probability flux
in y-direction \cite{GrHa71-2,Ha75}.  One finds, afterwards :
\begin{eqnarray}
  {p_{st}}\,\frac{d}{dH} (\alpha H -H^2)\,=\,D\,\frac{d}{dH} p_{st}\,.
\end{eqnarray}
It leads to Eq. (\ref{equ:SPDFVanDerPol}).

\section{Foundation of the algorithm}
\label{sec:found}
In this Appendix we show that the presented algorithm is described by a corresponding Master equation for the probability distribution
density $P_{i}(t)$ for the case of a Markovian hopping process
between boxes. We use a single index $i$ to label the boxes, but the argument applies to any spatial dimension $d$. We identify the dynamics of the stochastic system
which shall be simulated with the discrete stochastic dynamics of the
walkers. The latter is defined via the matrices of probabilities per
unit time $w(i\to i')$ which describe the hopping in the given
discretized space. We assume that it shall converge for sufficiently
small time scales and box lengths to the outgoing dynamics.

Let assume that we have the probability distribution density given at
time $t$ in every box with index $i$. It holds
\begin{eqnarray}
  \sum_{i} P_{i}(t) = 1.
\end{eqnarray}
We redistribute the probability in every box to $N$ walkers. In
result any walker $k=1,\dots,N$ of the same box gets an identical weight
\begin{eqnarray}
  q_{i}^k(t)\,=\,{P_{i}(t) \over N}\,.
\end{eqnarray}
until the next new redistribution. 

At later time $t+h$ the walkers may be still inside the box or may
have jumped to other boxes. Let $U_{i}$ denote all possible
box-indices which can be reached during a single step from the
${i}$-box. Then in accordance with the assumption above, the
probability is $w(i \to i') h$ that during $h$ a single walker
leaves ${i}$-box and jumps to the box with index $i'\,\in\, U_{i}$. By this hopping the walker $k$ carries the weight
$q_{i}^k(t)$ to the new box, which step is, obviously, connected with
a lost in the outgoing box. Therefore, the lost per particle for this
specific hopping from $i\to i'$ can be expressed as
\begin{eqnarray}
  w(i\to i')\,  h\, q_{i}^k(t).
\end{eqnarray}
The whole lost of weight will be realized on all possible hopping
channels. It is identical for all $N$ particles being located in the
present box. Hence, the full lost becomes 
\begin{align}
  N q_{i}^k(t) \sum_{i' \in  U_{i}}  w(i \to i') h =P_{i}(t)  \sum_{i' \in   U_{i}}  w(i \to i')\,  h\,.
\end{align}

Alternatively, one can also introduce $U^\prime_{i}$ as the boxes
where from walkers can reach the ${i}$-box. Then, gain of weight
transferred by every walker arriving at the $i$-box is
expressed, if $i' \in \, U^\prime_{i}$, by
\begin{eqnarray}
 w(i' \to i)\, h \, q_{i'}^k 
\end{eqnarray}
Again, summing over the different hopping steps and considering that all the
$N$ walkers inside the box with $i'\in  U^\prime_{i}$
reach the $i$ box, yields the gain 
\begin{align}
\sum_{i'\in  U^\prime_{i}}  w(i' \to i)  h  N q_{i'}^k(t) =\sum_{i'\in  U^\prime_{i}}  w(i' \to i) h P_{i'}(t) . 
\end{align}
Therefore, the balance of transported weight results in the following
shift of the full weight in the $i$-box at time $t+h$ compared
to the former one
\begin{eqnarray}
  \label{eq:master}
  P_{i}(t+h)\,-\,P_{i}(t)\,=&&\,- \, P_{i}(t) \, \sum_{i'\in  U_{i}}  w(i \to i')  h \nonumber\\
  &&+\sum_{i' \in  U^\prime_{i}}  w(i' \to i)\,  h P_{i'}(t)\, ,
\end{eqnarray}
plus corrections of order $O(h^2)$ corresponding to cases in which two or more walkers jump from one box to another during the time interval $h$.
After dividing by the time step $h$ and taking the limit $ h\to 0$ we
obtain the wanted Master equation for the discretized dynamics in the
boxed  phase space
\begin{eqnarray}
  \label{eq:master1}
 \partial_t P_{i}(t)\,=\, &-& \,P_{i}(t) \, \sum_{i'\in U_{i}}  w(i \to i')\, \nonumber\\
& +& \,\sum_{i' \in U^\prime_{i}}  w(i' \to i)\, P_{i'}(t)\, .
\end{eqnarray}

\bibliographystyle{apsrev4-1} 
\renewcommand{\bibname}{References} 


\begin{thebibliography}{42}%
\makeatletter
\providecommand \@ifxundefined [1]{%
 \@ifx{#1\undefined}
}%
\providecommand \@ifnum [1]{%
 \ifnum #1\expandafter \@firstoftwo
 \else \expandafter \@secondoftwo
 \fi
}%
\providecommand \@ifx [1]{%
 \ifx #1\expandafter \@firstoftwo
 \else \expandafter \@secondoftwo
 \fi
}%
\providecommand \natexlab [1]{#1}%
\providecommand \enquote  [1]{``#1''}%
\providecommand \bibnamefont  [1]{#1}%
\providecommand \bibfnamefont [1]{#1}%
\providecommand \citenamefont [1]{#1}%
\providecommand \href@noop [0]{\@secondoftwo}%
\providecommand \href [0]{\begingroup \@sanitize@url \@href}%
\providecommand \@href[1]{\@@startlink{#1}\@@href}%
\providecommand \@@href[1]{\endgroup#1\@@endlink}%
\providecommand \@sanitize@url [0]{\catcode `\\12\catcode `\$12\catcode
  `\&12\catcode `\#12\catcode `\^12\catcode `\_12\catcode `\%12\relax}%
\providecommand \@@startlink[1]{}%
\providecommand \@@endlink[0]{}%
\providecommand \url  [0]{\begingroup\@sanitize@url \@url }%
\providecommand \@url [1]{\endgroup\@href {#1}{\urlprefix }}%
\providecommand \urlprefix  [0]{URL }%
\providecommand \Eprint [0]{\href }%
\providecommand \doibase [0]{http://dx.doi.org/}%
\providecommand \selectlanguage [0]{\@gobble}%
\providecommand \bibinfo  [0]{\@secondoftwo}%
\providecommand \bibfield  [0]{\@secondoftwo}%
\providecommand \translation [1]{[#1]}%
\providecommand \BibitemOpen [0]{}%
\providecommand \bibitemStop [0]{}%
\providecommand \bibitemNoStop [0]{.\EOS\space}%
\providecommand \EOS [0]{\spacefactor3000\relax}%
\providecommand \BibitemShut  [1]{\csname bibitem#1\endcsname}%
\let\auto@bib@innerbib\@empty
\bibitem [{\citenamefont {Moss}\ and\ \citenamefont
  {McClintock}(1989)}]{moss1989noise}%
  \BibitemOpen
  \bibfield  {author} {\bibinfo {author} {\bibfnamefont {F.}~\bibnamefont
  {Moss}}\ and\ \bibinfo {author} {\bibfnamefont {P.~V.}\ \bibnamefont
  {McClintock}},\ }\href@noop {} {\emph {\bibinfo {title} {Noise in nonlinear
  dynamical systems}}},\ Vol.\ \bibinfo {volume} {I-III}\ (\bibinfo
  {publisher} {Cambridge University Press},\ \bibinfo {address} {New York},\
  \bibinfo {year} {1989})\BibitemShut {NoStop}%
\bibitem [{\citenamefont {G.~van Kampen}(1992)}]{van2004stochastic}%
  \BibitemOpen
  \bibfield  {author} {\bibinfo {author} {\bibfnamefont {N.}~\bibnamefont
  {G.~van Kampen}},\ }\href@noop {} {\emph {\bibinfo {title} {Stochastic
  processes in physics and chemistry}}}\ (\bibinfo  {publisher}
  {North-Holland},\ \bibinfo {address} {Amsterdam},\ \bibinfo {year}
  {1992})\BibitemShut {NoStop}%
\bibitem [{\citenamefont {Nicolis}\ and\ \citenamefont
  {Prigogine}(1977)}]{nicolis1977self}%
  \BibitemOpen
  \bibfield  {author} {\bibinfo {author} {\bibfnamefont {G.}~\bibnamefont
  {Nicolis}}\ and\ \bibinfo {author} {\bibfnamefont {I.}~\bibnamefont
  {Prigogine}},\ }\href@noop {} {\emph {\bibinfo {title} {Self-Organization in
  Nonequilibrium Systems: From Dissipative Structures to Order Through
  Fluctuations}}}\ (\bibinfo  {publisher} {Wiley-Interscience},\ \bibinfo
  {address} {New York},\ \bibinfo {year} {1977})\BibitemShut {NoStop}%
\bibitem [{\citenamefont {Mannella}\ and\ \citenamefont
  {Palleschi}(1989)}]{mannella1989fast}%
  \BibitemOpen
  \bibfield  {author} {\bibinfo {author} {\bibfnamefont {R.}~\bibnamefont
  {Mannella}}\ and\ \bibinfo {author} {\bibfnamefont {V.}~\bibnamefont
  {Palleschi}},\ }\href@noop {} {\bibfield  {journal} {\bibinfo  {journal}
  {Phys. Rev. A}\ }\textbf {\bibinfo {volume} {40}},\ \bibinfo {pages} {3381}
  (\bibinfo {year} {1989})}\BibitemShut {NoStop}%
\bibitem [{\citenamefont {Mannella}(2000)}]{mannella2000lecture}%
  \BibitemOpen
  \bibfield  {author} {\bibinfo {author} {\bibfnamefont {R.}~\bibnamefont
  {Mannella}},\ }in\ \href@noop {} {\emph {\bibinfo {booktitle} {Stochastic
  Processes in Physics, Chemistry, and Biology}}},\ \bibinfo {series} {Lecture
  Notes in Physics}, Vol.\ \bibinfo {volume} {557},\ \bibinfo {editor} {edited
  by\ \bibinfo {editor} {\bibfnamefont {J.~A.}\ \bibnamefont {Freund}}\ and\
  \bibinfo {editor} {\bibfnamefont {T.}~\bibnamefont {P{\"o}schel}}}\ (\bibinfo
   {publisher} {Springer},\ \bibinfo {address} {Berlin},\ \bibinfo {year}
  {2000})\ p.\ \bibinfo {pages} {353}\BibitemShut {NoStop}%
\bibitem [{\citenamefont {Burrage}(1999)}]{burrage1999runge}%
  \BibitemOpen
  \bibfield  {author} {\bibinfo {author} {\bibfnamefont {P.}~\bibnamefont
  {Burrage}},\ }\emph {\bibinfo {title} {Runge-Kutta methods for stochastic
  differential equations}},\ \href@noop {} {Ph.D. thesis},\ \bibinfo  {school}
  {University of Queensland} (\bibinfo {year} {1999})\BibitemShut {NoStop}%
\bibitem [{\citenamefont {Kloeden}\ and\ \citenamefont
  {Platen}(2011)}]{kloeden2011numerical}%
  \BibitemOpen
  \bibfield  {author} {\bibinfo {author} {\bibfnamefont {P.}~\bibnamefont
  {Kloeden}}\ and\ \bibinfo {author} {\bibfnamefont {E.}~\bibnamefont
  {Platen}},\ }\href@noop {} {\emph {\bibinfo {title} {Numerical solution of
  stochastic differential equations}}},\ Vol.~\bibinfo {volume} {23}\ (\bibinfo
   {publisher} {Springer},\ \bibinfo {address} {Berlin},\ \bibinfo {year}
  {2011})\BibitemShut {NoStop}%
\bibitem [{\citenamefont {Bhanot}\ \emph {et~al.}(1987)\citenamefont {Bhanot},
  \citenamefont {Salvador}, \citenamefont {Black}, \citenamefont {Carter},\
  and\ \citenamefont {Toral}}]{bhanot1987accurate}%
  \BibitemOpen
  \bibfield  {author} {\bibinfo {author} {\bibfnamefont {G.}~\bibnamefont
  {Bhanot}}, \bibinfo {author} {\bibfnamefont {R.}~\bibnamefont {Salvador}},
  \bibinfo {author} {\bibfnamefont {S.}~\bibnamefont {Black}}, \bibinfo
  {author} {\bibfnamefont {P.}~\bibnamefont {Carter}}, \ and\ \bibinfo {author}
  {\bibfnamefont {R.}~\bibnamefont {Toral}},\ }\href@noop {} {\bibfield
  {journal} {\bibinfo  {journal} {Phys. Rev. Lett.}\ }\textbf {\bibinfo
  {volume} {59}},\ \bibinfo {pages} {803} (\bibinfo {year} {1987})}\BibitemShut
  {NoStop}%
\bibitem [{\citenamefont {Giardina}\ \emph {et~al.}(2006)\citenamefont
  {Giardina}, \citenamefont {Kurchan},\ and\ \citenamefont
  {Peliti}}]{giardina2006direct}%
  \BibitemOpen
  \bibfield  {author} {\bibinfo {author} {\bibfnamefont {C.}~\bibnamefont
  {Giardina}}, \bibinfo {author} {\bibfnamefont {J.}~\bibnamefont {Kurchan}}, \
  and\ \bibinfo {author} {\bibfnamefont {L.}~\bibnamefont {Peliti}},\
  }\href@noop {} {\bibfield  {journal} {\bibinfo  {journal} {Phys. Rev. Lett.}\
  }\textbf {\bibinfo {volume} {96}},\ \bibinfo {pages} {120603} (\bibinfo
  {year} {2006})}\BibitemShut {NoStop}%
\bibitem [{\citenamefont {Dickson}\ and\ \citenamefont
  {Dinner}(2010)}]{dickson2010enhanced}%
  \BibitemOpen
  \bibfield  {author} {\bibinfo {author} {\bibfnamefont {A.}~\bibnamefont
  {Dickson}}\ and\ \bibinfo {author} {\bibfnamefont {A.}~\bibnamefont
  {Dinner}},\ }\href@noop {} {\bibfield  {journal} {\bibinfo  {journal} {Annu.
  Rev. Phys. Chem.}\ }\textbf {\bibinfo {volume} {61}},\ \bibinfo {pages} {441}
  (\bibinfo {year} {2010})}\BibitemShut {NoStop}%
\bibitem [{\citenamefont {Wang}\ and\ \citenamefont
  {P.~Landau}(2001)}]{wang2001efficient}%
  \BibitemOpen
  \bibfield  {author} {\bibinfo {author} {\bibfnamefont {F.}~\bibnamefont
  {Wang}}\ and\ \bibinfo {author} {\bibfnamefont {D.}~\bibnamefont
  {P.~Landau}},\ }\href@noop {} {\bibfield  {journal} {\bibinfo  {journal}
  {Phys. Rev. Lett.}\ }\textbf {\bibinfo {volume} {86}},\ \bibinfo {pages}
  {2050} (\bibinfo {year} {2001})}\BibitemShut {NoStop}%
\bibitem [{\citenamefont {Torrie}\ and\ \citenamefont
  {Valleau}(1977)}]{torrie1977nonphysical}%
  \BibitemOpen
  \bibfield  {author} {\bibinfo {author} {\bibfnamefont {G.}~\bibnamefont
  {Torrie}}\ and\ \bibinfo {author} {\bibfnamefont {J.}~\bibnamefont
  {Valleau}},\ }\href@noop {} {\bibfield  {journal} {\bibinfo  {journal} {J.
  Comp. Phys.}\ }\textbf {\bibinfo {volume} {23}},\ \bibinfo {pages} {187}
  (\bibinfo {year} {1977})}\BibitemShut {NoStop}%
\bibitem [{\citenamefont {Warmflash}\ \emph {et~al.}(2007)\citenamefont
  {Warmflash}, \citenamefont {Bhimalapuram},\ and\ \citenamefont
  {Dinner}}]{warmflash2007umbrella}%
  \BibitemOpen
  \bibfield  {author} {\bibinfo {author} {\bibfnamefont {A.}~\bibnamefont
  {Warmflash}}, \bibinfo {author} {\bibfnamefont {P.}~\bibnamefont
  {Bhimalapuram}}, \ and\ \bibinfo {author} {\bibfnamefont {A.}~\bibnamefont
  {Dinner}},\ }\href@noop {} {\bibfield  {journal} {\bibinfo  {journal} {J.
  Chem. Phys.}\ }\textbf {\bibinfo {volume} {127}},\ \bibinfo {pages} {154112}
  (\bibinfo {year} {2007})}\BibitemShut {NoStop}%
\bibitem [{\citenamefont {Allen}\ \emph {et~al.}(2009)\citenamefont {Allen},
  \citenamefont {Valeriani},\ and\ \citenamefont {ten
  Wolde}}]{allen2009forward}%
  \BibitemOpen
  \bibfield  {author} {\bibinfo {author} {\bibfnamefont {R.}~\bibnamefont
  {Allen}}, \bibinfo {author} {\bibfnamefont {C.}~\bibnamefont {Valeriani}}, \
  and\ \bibinfo {author} {\bibfnamefont {P.}~\bibnamefont {ten Wolde}},\
  }\href@noop {} {\bibfield  {journal} {\bibinfo  {journal} {J. Phys.: Condens.
  Matter}\ }\textbf {\bibinfo {volume} {21}},\ \bibinfo {pages} {463102}
  (\bibinfo {year} {2009})}\BibitemShut {NoStop}%
\bibitem [{\citenamefont {Dickson}\ \emph {et~al.}(2009)\citenamefont
  {Dickson}, \citenamefont {Warmflash},\ and\ \citenamefont
  {Dinner}}]{dickson2009separating}%
  \BibitemOpen
  \bibfield  {author} {\bibinfo {author} {\bibfnamefont {A.}~\bibnamefont
  {Dickson}}, \bibinfo {author} {\bibfnamefont {A.}~\bibnamefont {Warmflash}},
  \ and\ \bibinfo {author} {\bibfnamefont {A.}~\bibnamefont {Dinner}},\
  }\href@noop {} {\bibfield  {journal} {\bibinfo  {journal} {J. Chem. Phys.}\
  }\textbf {\bibinfo {volume} {131}},\ \bibinfo {pages} {154104} (\bibinfo
  {year} {2009})}\BibitemShut {NoStop}%
\bibitem [{\citenamefont {Valeriani}\ \emph {et~al.}(2007)\citenamefont
  {Valeriani}, \citenamefont {Allen}, \citenamefont {Morelli}, \citenamefont
  {Frenkel},\ and\ \citenamefont {Wolde}}]{valeriani2009computing}%
  \BibitemOpen
  \bibfield  {author} {\bibinfo {author} {\bibfnamefont {C.}~\bibnamefont
  {Valeriani}}, \bibinfo {author} {\bibfnamefont {R.}~\bibnamefont {Allen}},
  \bibinfo {author} {\bibfnamefont {M.}~\bibnamefont {Morelli}}, \bibinfo
  {author} {\bibfnamefont {D.}~\bibnamefont {Frenkel}}, \ and\ \bibinfo
  {author} {\bibfnamefont {P.}~\bibnamefont {Wolde}},\ }\href@noop {}
  {\bibfield  {journal} {\bibinfo  {journal} {J. Chem. Phys.}\ }\textbf
  {\bibinfo {volume} {127}},\ \bibinfo {pages} {114109} (\bibinfo {year}
  {2007})}\BibitemShut {NoStop}%
\bibitem [{\citenamefont {Huber}\ and\ \citenamefont
  {Kim}(1996)}]{huber1996weighted}%
  \BibitemOpen
  \bibfield  {author} {\bibinfo {author} {\bibfnamefont {G.}~\bibnamefont
  {Huber}}\ and\ \bibinfo {author} {\bibfnamefont {S.}~\bibnamefont {Kim}},\
  }\href@noop {} {\bibfield  {journal} {\bibinfo  {journal} {Biophys. J.}\
  }\textbf {\bibinfo {volume} {70}},\ \bibinfo {pages} {97} (\bibinfo {year}
  {1996})}\BibitemShut {NoStop}%
\bibitem [{\citenamefont {Bhatt}\ \emph {et~al.}(2010)\citenamefont {Bhatt},
  \citenamefont {Zhang},\ and\ \citenamefont {Zuckerman}}]{bhatt2010steady}%
  \BibitemOpen
  \bibfield  {author} {\bibinfo {author} {\bibfnamefont {D.}~\bibnamefont
  {Bhatt}}, \bibinfo {author} {\bibfnamefont {B.}~\bibnamefont {Zhang}}, \ and\
  \bibinfo {author} {\bibfnamefont {D.}~\bibnamefont {Zuckerman}},\ }\href@noop
  {} {\bibfield  {journal} {\bibinfo  {journal} {J. Chem. Phys.}\ }\textbf
  {\bibinfo {volume} {133}},\ \bibinfo {pages} {014110} (\bibinfo {year}
  {2010})}\BibitemShut {NoStop}%
\bibitem [{\citenamefont {Zhang}\ \emph {et~al.}(2010)\citenamefont {Zhang},
  \citenamefont {Jasnow},\ and\ \citenamefont {Zuckerman}}]{zhang2010weighted}%
  \BibitemOpen
  \bibfield  {author} {\bibinfo {author} {\bibfnamefont {B.}~\bibnamefont
  {Zhang}}, \bibinfo {author} {\bibfnamefont {D.}~\bibnamefont {Jasnow}}, \
  and\ \bibinfo {author} {\bibfnamefont {D.}~\bibnamefont {Zuckerman}},\
  }\href@noop {} {\bibfield  {journal} {\bibinfo  {journal} {J. Chem. Phys.}\
  }\textbf {\bibinfo {volume} {132}},\ \bibinfo {pages} {054107} (\bibinfo
  {year} {2010})}\BibitemShut {NoStop}%
\bibitem [{\citenamefont {Bhatt}\ and\ \citenamefont
  {Bahar}(2012)}]{bhatt2012adaptive}%
  \BibitemOpen
  \bibfield  {author} {\bibinfo {author} {\bibfnamefont {D.}~\bibnamefont
  {Bhatt}}\ and\ \bibinfo {author} {\bibfnamefont {I.}~\bibnamefont {Bahar}},\
  }\href@noop {} {\bibfield  {journal} {\bibinfo  {journal} {J. Chem. Phys.}\
  }\textbf {\bibinfo {volume} {137}},\ \bibinfo {pages} {104101} (\bibinfo
  {year} {2012})}\BibitemShut {NoStop}%
\bibitem [{\citenamefont {Bhatt}\ and\ \citenamefont
  {Zuckerman}(2011)}]{bhatt2011beyond}%
  \BibitemOpen
  \bibfield  {author} {\bibinfo {author} {\bibfnamefont {D.}~\bibnamefont
  {Bhatt}}\ and\ \bibinfo {author} {\bibfnamefont {D.}~\bibnamefont
  {Zuckerman}},\ }\href@noop {} {\bibfield  {journal} {\bibinfo  {journal} {J.
  Chem. Theory Comput.}\ }\textbf {\bibinfo {volume} {7}},\ \bibinfo {pages}
  {2520} (\bibinfo {year} {2011})}\BibitemShut {NoStop}%
\bibitem [{\citenamefont {Graham}\ and\ \citenamefont
  {Haken}(1971{\natexlab{a}})}]{GrHa71-2}%
  \BibitemOpen
  \bibfield  {author} {\bibinfo {author} {\bibfnamefont {R.}~\bibnamefont
  {Graham}}\ and\ \bibinfo {author} {\bibfnamefont {H.}~\bibnamefont {Haken}},\
  }\href@noop {} {\bibfield  {journal} {\bibinfo  {journal} {Z. Phys.}\
  }\textbf {\bibinfo {volume} {243}},\ \bibinfo {pages} {289} (\bibinfo {year}
  {1971}{\natexlab{a}})}\BibitemShut {NoStop}%
\bibitem [{\citenamefont {Graham}\ and\ \citenamefont
  {Haken}(1971{\natexlab{b}})}]{GrHa71}%
  \BibitemOpen
  \bibfield  {author} {\bibinfo {author} {\bibfnamefont {R.}~\bibnamefont
  {Graham}}\ and\ \bibinfo {author} {\bibfnamefont {H.}~\bibnamefont {Haken}},\
  }\href@noop {} {\bibfield  {journal} {\bibinfo  {journal} {Z. Phys.}\
  }\textbf {\bibinfo {volume} {245}},\ \bibinfo {pages} {141} (\bibinfo {year}
  {1971}{\natexlab{b}})}\BibitemShut {NoStop}%
\bibitem [{\citenamefont {Haken}(1975)}]{Ha75}%
  \BibitemOpen
  \bibfield  {author} {\bibinfo {author} {\bibfnamefont {H.}~\bibnamefont
  {Haken}},\ }\href@noop {} {\bibfield  {journal} {\bibinfo  {journal} {Rev.
  Mod. Phys.}\ }\textbf {\bibinfo {volume} {47}},\ \bibinfo {pages} {67}
  (\bibinfo {year} {1975})}\BibitemShut {NoStop}%
\bibitem [{\citenamefont {Risken}(1984)}]{risken}%
  \BibitemOpen
  \bibfield  {author} {\bibinfo {author} {\bibfnamefont {H.}~\bibnamefont
  {Risken}},\ }\href@noop {} {\emph {\bibinfo {title} {The Fokker Planck
  equation}}},\ Vol.~\bibinfo {volume} {23}\ (\bibinfo  {publisher}
  {Springer},\ \bibinfo {address} {Berlin},\ \bibinfo {year}
  {1984})\BibitemShut {NoStop}%
\bibitem [{\citenamefont {Schimansky-Geier}\ \emph {et~al.}(1985)\citenamefont
  {Schimansky-Geier}, \citenamefont {Tolstopjatenko},\ and\ \citenamefont
  {Ebeling}}]{LsgTol85}%
  \BibitemOpen
  \bibfield  {author} {\bibinfo {author} {\bibfnamefont {L.}~\bibnamefont
  {Schimansky-Geier}}, \bibinfo {author} {\bibfnamefont {A.}~\bibnamefont
  {Tolstopjatenko}}, \ and\ \bibinfo {author} {\bibfnamefont {W.}~\bibnamefont
  {Ebeling}},\ }\href@noop {} {\bibfield  {journal} {\bibinfo  {journal} {Phys.
  Lett.}\ }\textbf {\bibinfo {volume} {108A}},\ \bibinfo {pages} {329}
  (\bibinfo {year} {1985})}\BibitemShut {NoStop}%
\bibitem [{\citenamefont {Ebeling}\ and\ \citenamefont
  {Engel-Herbert}(1980)}]{ebeling1980influence}%
  \BibitemOpen
  \bibfield  {author} {\bibinfo {author} {\bibfnamefont {W.}~\bibnamefont
  {Ebeling}}\ and\ \bibinfo {author} {\bibfnamefont {H.}~\bibnamefont
  {Engel-Herbert}},\ }\href@noop {} {\bibfield  {journal} {\bibinfo  {journal}
  {Physica A}\ }\textbf {\bibinfo {volume} {104}},\ \bibinfo {pages} {378}
  (\bibinfo {year} {1980})}\BibitemShut {NoStop}%
\bibitem [{\citenamefont {Klimontovich}(1980)}]{klimo}%
  \BibitemOpen
  \bibfield  {author} {\bibinfo {author} {\bibfnamefont {Y.}~\bibnamefont
  {Klimontovich}},\ }\href@noop {} {\emph {\bibinfo {title} {Kinetic theory of
  elecromagnetic processes}}}\ (\bibinfo  {publisher} {Springer},\ \bibinfo
  {address} {Berlin},\ \bibinfo {year} {1983 (in Russian: Nauka, Moscow,
  1980)})\BibitemShut {NoStop}%
\bibitem [{\citenamefont {Gilsing}\ and\ \citenamefont
  {Shardlow}(2007)}]{gilsing2007sdelab}%
  \BibitemOpen
  \bibfield  {author} {\bibinfo {author} {\bibfnamefont {H.}~\bibnamefont
  {Gilsing}}\ and\ \bibinfo {author} {\bibfnamefont {T.}~\bibnamefont
  {Shardlow}},\ }\href@noop {} {\bibfield  {journal} {\bibinfo  {journal} {J.
  Comput. Appl. Math.}\ }\textbf {\bibinfo {volume} {205}},\ \bibinfo {pages}
  {1002} (\bibinfo {year} {2007})}\BibitemShut {NoStop}%
\bibitem [{\citenamefont {Gammaitoni}\ \emph {et~al.}(1998)\citenamefont
  {Gammaitoni}, \citenamefont {H{\"a}nggi}, \citenamefont {Jung},\ and\
  \citenamefont {Marchesoni}}]{gammaitoni1998stochastic}%
  \BibitemOpen
  \bibfield  {author} {\bibinfo {author} {\bibfnamefont {L.}~\bibnamefont
  {Gammaitoni}}, \bibinfo {author} {\bibfnamefont {P.}~\bibnamefont
  {H{\"a}nggi}}, \bibinfo {author} {\bibfnamefont {P.}~\bibnamefont {Jung}}, \
  and\ \bibinfo {author} {\bibfnamefont {F.}~\bibnamefont {Marchesoni}},\
  }\href@noop {} {\bibfield  {journal} {\bibinfo  {journal} {Rev. Mod. Phys.}\
  }\textbf {\bibinfo {volume} {70}},\ \bibinfo {pages} {223} (\bibinfo {year}
  {1998})}\BibitemShut {NoStop}%
\bibitem [{\citenamefont {H{\"a}nggi}\ \emph {et~al.}(1990)\citenamefont
  {H{\"a}nggi}, \citenamefont {P.},\ and\ \citenamefont
  {Borkovec}}]{hanggi_rmp90}%
  \BibitemOpen
  \bibfield  {author} {\bibinfo {author} {\bibfnamefont {P.}~\bibnamefont
  {H{\"a}nggi}}, \bibinfo {author} {\bibfnamefont {T.}~\bibnamefont {P.}}, \
  and\ \bibinfo {author} {\bibfnamefont {N.}~\bibnamefont {Borkovec}},\
  }\href@noop {} {\bibfield  {journal} {\bibinfo  {journal} {Rev. Mod. Phys.}\
  }\textbf {\bibinfo {volume} {62}},\ \bibinfo {pages} {251} (\bibinfo {year}
  {1990})}\BibitemShut {NoStop}%
\bibitem [{\citenamefont {Tuckwell}(1988)}]{tuckwell}%
  \BibitemOpen
  \bibfield  {author} {\bibinfo {author} {\bibfnamefont {H.}~\bibnamefont
  {Tuckwell}},\ }\href@noop {} {\emph {\bibinfo {title} {Introduction to
  theoretical neurobiology}}},\ Vol.~\bibinfo {volume} {2}\ (\bibinfo
  {publisher} {Cambridge University Press},\ \bibinfo {address} {New York},\
  \bibinfo {year} {1988})\BibitemShut {NoStop}%
\bibitem [{\citenamefont {Ebeling}\ \emph {et~al.}(1986)\citenamefont
  {Ebeling}, \citenamefont {Herzel}, \citenamefont {Richert},\ and\
  \citenamefont {Schimansky-Geier}}]{EbHe86}%
  \BibitemOpen
  \bibfield  {author} {\bibinfo {author} {\bibfnamefont {W.}~\bibnamefont
  {Ebeling}}, \bibinfo {author} {\bibfnamefont {H.}~\bibnamefont {Herzel}},
  \bibinfo {author} {\bibfnamefont {W.}~\bibnamefont {Richert}}, \ and\
  \bibinfo {author} {\bibfnamefont {L.}~\bibnamefont {Schimansky-Geier}},\
  }\href@noop {} {\bibfield  {journal} {\bibinfo  {journal} {Zeitschrift f.
  angew. Math. und Mech.}\ }\textbf {\bibinfo {volume} {66}},\ \bibinfo {pages}
  {141} (\bibinfo {year} {1986})}\BibitemShut {NoStop}%
\bibitem [{\citenamefont {Lekkas}\ \emph {et~al.}(1988)\citenamefont {Lekkas},
  \citenamefont {Schimansky-Geier},\ and\ \citenamefont
  {Engel-Herbert}}]{LekLsg88}%
  \BibitemOpen
  \bibfield  {author} {\bibinfo {author} {\bibfnamefont {K.}~\bibnamefont
  {Lekkas}}, \bibinfo {author} {\bibfnamefont {L.}~\bibnamefont
  {Schimansky-Geier}}, \ and\ \bibinfo {author} {\bibfnamefont
  {H.}~\bibnamefont {Engel-Herbert}},\ }\href@noop {} {\bibfield  {journal}
  {\bibinfo  {journal} {Z. Phys. B - Condensed Matter}\ }\textbf {\bibinfo
  {volume} {70}},\ \bibinfo {pages} {517} (\bibinfo {year} {1988})}\BibitemShut
  {NoStop}%
\bibitem [{\citenamefont {Debnath}\ \emph {et~al.}(1990)\citenamefont
  {Debnath}, \citenamefont {Moss}, \citenamefont {Leiber}, \citenamefont
  {Risken},\ and\ \citenamefont {Marchesoni}}]{debnath1990holes}%
  \BibitemOpen
  \bibfield  {author} {\bibinfo {author} {\bibfnamefont {G.}~\bibnamefont
  {Debnath}}, \bibinfo {author} {\bibfnamefont {F.}~\bibnamefont {Moss}},
  \bibinfo {author} {\bibfnamefont {T.}~\bibnamefont {Leiber}}, \bibinfo
  {author} {\bibfnamefont {H.}~\bibnamefont {Risken}}, \ and\ \bibinfo {author}
  {\bibfnamefont {F.}~\bibnamefont {Marchesoni}},\ }\href@noop {} {\bibfield
  {journal} {\bibinfo  {journal} {Phys. Rev. A}\ }\textbf {\bibinfo {volume}
  {42}},\ \bibinfo {pages} {703} (\bibinfo {year} {1990})}\BibitemShut
  {NoStop}%
\bibitem [{\citenamefont {H{\"a}nggi}\ and\ \citenamefont
  {Jung}(1995)}]{HaJu95}%
  \BibitemOpen
  \bibfield  {author} {\bibinfo {author} {\bibfnamefont {P.}~\bibnamefont
  {H{\"a}nggi}}\ and\ \bibinfo {author} {\bibfnamefont {P.}~\bibnamefont
  {Jung}},\ }\href@noop {} {\bibfield  {journal} {\bibinfo  {journal} {Adv.
  Chem. Phys.}\ }\textbf {\bibinfo {volume} {89}},\ \bibinfo {pages} {239}
  (\bibinfo {year} {1995})}\BibitemShut {NoStop}%
\bibitem [{\citenamefont {Fitzhugh}(1961)}]{fitzhugh1961impulses}%
  \BibitemOpen
  \bibfield  {author} {\bibinfo {author} {\bibfnamefont {R.}~\bibnamefont
  {Fitzhugh}},\ }\href@noop {} {\bibfield  {journal} {\bibinfo  {journal}
  {Biophys. J.}\ }\textbf {\bibinfo {volume} {1}},\ \bibinfo {pages} {445}
  (\bibinfo {year} {1961})}\BibitemShut {NoStop}%
\bibitem [{\citenamefont {Lindner}\ \emph {et~al.}(2004)\citenamefont
  {Lindner}, \citenamefont {Garc{\i}a-Ojalvo}, \citenamefont {Neiman},\ and\
  \citenamefont {Schimansky-Geier}}]{lindner2004effects}%
  \BibitemOpen
  \bibfield  {author} {\bibinfo {author} {\bibfnamefont {B.}~\bibnamefont
  {Lindner}}, \bibinfo {author} {\bibfnamefont {J.}~\bibnamefont
  {Garc{\i}a-Ojalvo}}, \bibinfo {author} {\bibfnamefont {A.}~\bibnamefont
  {Neiman}}, \ and\ \bibinfo {author} {\bibfnamefont {L.}~\bibnamefont
  {Schimansky-Geier}},\ }\href@noop {} {\bibfield  {journal} {\bibinfo
  {journal} {Phys. Rept.}\ }\textbf {\bibinfo {volume} {392}},\ \bibinfo
  {pages} {321} (\bibinfo {year} {2004})}\BibitemShut {NoStop}%
\bibitem [{\citenamefont {Gunton}\ \emph {et~al.}(2003)\citenamefont {Gunton},
  \citenamefont {Toral}, \citenamefont {Mirasso}, \citenamefont {Gracheva}
  \emph {et~al.}}]{gunton2003role}%
  \BibitemOpen
  \bibfield  {author} {\bibinfo {author} {\bibfnamefont {J.}~\bibnamefont
  {Gunton}}, \bibinfo {author} {\bibfnamefont {R.}~\bibnamefont {Toral}},
  \bibinfo {author} {\bibfnamefont {C.}~\bibnamefont {Mirasso}}, \bibinfo
  {author} {\bibfnamefont {M.}~\bibnamefont {Gracheva}},  \emph {et~al.},\
  }\href@noop {} {\bibfield  {journal} {\bibinfo  {journal} {Recent. Res.
  Devel. Applied Phys.}\ }\textbf {\bibinfo {volume} {6}},\ \bibinfo {pages}
  {497} (\bibinfo {year} {2003})}\BibitemShut {NoStop}%
\bibitem [{\citenamefont {Toral}\ \emph {et~al.}(2003)\citenamefont {Toral},
  \citenamefont {Masoller}, \citenamefont {Mirasso}, \citenamefont {Ciszak},\
  and\ \citenamefont {Calvo}}]{toral2003characterization}%
  \BibitemOpen
  \bibfield  {author} {\bibinfo {author} {\bibfnamefont {R.}~\bibnamefont
  {Toral}}, \bibinfo {author} {\bibfnamefont {C.}~\bibnamefont {Masoller}},
  \bibinfo {author} {\bibfnamefont {C.}~\bibnamefont {Mirasso}}, \bibinfo
  {author} {\bibfnamefont {M.}~\bibnamefont {Ciszak}}, \ and\ \bibinfo {author}
  {\bibfnamefont {O.}~\bibnamefont {Calvo}},\ }\href@noop {} {\bibfield
  {journal} {\bibinfo  {journal} {Physica A}\ }\textbf {\bibinfo {volume}
  {325}},\ \bibinfo {pages} {192} (\bibinfo {year} {2003})}\BibitemShut
  {NoStop}%
\bibitem [{\citenamefont {Toral}\ \emph {et~al.}(2007)\citenamefont {Toral},
  \citenamefont {Mirasso},\ and\ \citenamefont {Gunton}}]{toral2007system}%
  \BibitemOpen
  \bibfield  {author} {\bibinfo {author} {\bibfnamefont {R.}~\bibnamefont
  {Toral}}, \bibinfo {author} {\bibfnamefont {C.}~\bibnamefont {Mirasso}}, \
  and\ \bibinfo {author} {\bibfnamefont {J.}~\bibnamefont {Gunton}},\
  }\href@noop {} {\bibfield  {journal} {\bibinfo  {journal} {Europhys. Lett.}\
  }\textbf {\bibinfo {volume} {61}},\ \bibinfo {pages} {162} (\bibinfo {year}
  {2007})}\BibitemShut {NoStop}%
\bibitem [{\citenamefont {Kostur}\ \emph {et~al.}(2003)\citenamefont {Kostur},
  \citenamefont {Sailer},\ and\ \citenamefont
  {Schimansky-Geier}}]{kostur2003stationary}%
  \BibitemOpen
  \bibfield  {author} {\bibinfo {author} {\bibfnamefont {M.}~\bibnamefont
  {Kostur}}, \bibinfo {author} {\bibfnamefont {X.}~\bibnamefont {Sailer}}, \
  and\ \bibinfo {author} {\bibfnamefont {L.}~\bibnamefont {Schimansky-Geier}},\
  }\href@noop {} {\bibfield  {journal} {\bibinfo  {journal} {Fluct. Noise
  Lett.}\ }\textbf {\bibinfo {volume} {3}},\ \bibinfo {pages} {155} (\bibinfo
  {year} {2003})}\BibitemShut {NoStop}%
\end{thebibliography}

%

\end{document}